\documentclass[12pt]{article}
\usepackage{amssymb}
\def\thebibliography#1{\section*{References}\list
 {}{\setlength\labelwidth{1.4em}\leftmargin\labelwidth
 \setlength\parsep{0pt}\setlength\itemsep{0pt}
 \setlength{\itemindent}{-\leftmargin}
 \usecounter{enumi}}}

\begin{document}
\begin{center}
{\bf Giant Radio Galaxies as a probe of the cosmological evolution of the
IGM, I. Preliminary deep detections and low-resolution spectroscopy with the SALT}
\end{center}

\vspace{2mm}
\begin{center}
J. Machalski$^1$, D. Kozie{\l}-Wierzbowska$^1$, M. Jamrozy$^1$\\
$^1$ Obserwatorium Astronomiczne, Uniwersytet Jagiello{\'n}ski,\\ ul. Orla 171, 30-244, Krak{\'o}w, Poland\\
(machalsk@oa.uj.edu.pl, eddie@oa.uj.edu.pl, jamrozy@oa.uj.edu.pl)
\vspace{3mm}
{\bf Abstract}
\end{center}
\vspace{3mm}
A problem of the cosmological evolution of the IGM is recalled and a necessity
to find distant ($z>0.5$) ``giant'' radio galaxies (GRGs) with the lobe energy
densities lower than about $10^{-14}$ J\,m$^{-3}$ to solve this problem is
emphasized. Therefore we undertake a search for such GRGs on the southern sky
hemisphere using the SALT. In this paper we present a selected sample of the
GRG candidates and the first deep detections of distant host galaxies, as well
as the low-resolution spectra of the galaxies identified on the DSS frames. The
data collected during the {\sl Performance Verification} (P--V) phase show that
21 of 35 galaxies with the spectroscopic redshift have the projected linear
size greater than 1 Mpc (for $H_{0}$=71 km\,s$^{-1}$Mpc$^{-1}$). However their
redshifts do not exceed the value of 0.4 and the energy density in only two of
them is less than $10^{-14}$ J\,m$^{-3}$. A photometric redshift estimate of
one of them (J1420$-$0545) suggests a linear extent larger than 4.8 Mpc, i.e.
a larger than that of 3C236, the largest GRG known up to now.


\vspace{4mm}
\noindent
\underline{1. Introduction}

In the adiabatically expanding Universe filled with a hot, diffuse and uniform
IGM, the IGM pressure should increase with redshift as $p_{\rm IGM}(z)=p_{0}(1+z)^{5}$,
where $p_{0}$ is the present-day pressure (cf. Subrahmanyan \& Saripalli 1993).
Various methods have been employed in determining the density and pressure of the
IGM. The most effective are X-ray observations of the hot gas around radio sources
of Fanaroff-Riley type II (FRII) yielding direct estimates of the gas
density (Crawford et al. 1999; Hardcastle et al. 2002; Croston et al. 2004, 2005).
Assuming an equation of state of this gas, a value of $p_{\rm IGM}$ can be specified.
However, both the AGN and the large-scale radio structure likely contribute to the
X-ray emission (cf. Kaiser \& Alexander 1999), thus the properties of the environment
may be considerably influenced by the presence of the radio source itself.  
As the IGM forms the environment of the radio source, the internal physical
properties of its diffuse lobes  were expected to reflect directly $p_{IGM}(z)$.
However, the cosmic background measurements (e.g. with COBE) ruled out an IGM
with the pressure indicated by typical radio structures. This suggested that these
regions were not thermally confined. Indeed, the very existence of apparent hot
spots at the edges of double FRII-type radio sources indicates that the jets ejected
from the AGN encounter resistance to their propagation, and these advancing hot spots
are likely confined by ram-pressure of the IGM. But the more tenuous material, outside
the jets, forming a bridge between the nucleus and the radio lobes is expected to have
attained an equilibrium state in which the energy density of the particles and magnetic
field is balanced, and the pressure of the relativistic plasma equals the pressure of
the gaseous environment. Consequently, a determination of physical conditions in these
diffuse bridges of large-size double radio sources (being away from the ram-pressure
confined the source heads) gives an independent tool of probing the pressure of the IGM.

This physical conditions, specifically the internal pressure, can be determined from
the dynamical considerations. In the analytical model of the source dynamics of
Kaiser \& Alexander (1997) and its further modifications (e.g. Blundell et al. 1999)
the pressure is dependent on the gas density, power of the jet, and age of the radio
structure. As a certain determination of these parameters is not easy, the internal
pressure is usually estimated from the minimum energy condition corresponding to an 
equipartition of energy between the relativistic particles and the magnetic field.  
The minimum energy can be calculated from the total luminosity and the size (volume)
of a given radio source (e.g. Pacholczyk 1970; Miley 1980).
The largest extragalactic radio sources (D$>$1 Mpc, frequently
called {\sc Giants} or {\sc Giant Radio Galaxies} [GRGs] though a small fraction
of them are classified as QSOs), due to their very extended radio lobes and sometime
bridges linking the opposite lobes, offer a unique
probe the intergalactic and intracluster environment directly on Mpc scales
allowing measurements of the IGM structure, pressure and their evolution out to
high redshifts (cf. Strom \& Willis 1980; Subrahmanyan \& Saripalli 1993; Mack et
al. 1998).  

Subrahmanyan \& Saripalli (1993) contended that the lower and upper limit values of the
present-day IGM pressure are 0.5 and 2$\times$10$^{-15}$ N\,m$^{-2}$, respectively, but
further investigations raised those values at least twice (e.g. Ishwara-Chandra \&
Saikia (1999). The problem of the cosmic pressure evolution was also considered by
Schoenmakers et al. (2000) who found no evidence for such a strong evolution up to
the redshift of about 0.3. They concluded the apparent pressure increase with redshift
can be explained by two effects: the use of flux-limited samples and the method by which
the lobe pressure is calculated. The latter aspect is discussed more in details in Section~6.
Therefore, in order to verify the hypothesis about the IGM pressure
evolution concordant with $(1+z)^{5}$, it is necessary to find GRGs with the lobe
energy densities of $<10^{-14}$ J\,m$^{-3}$ at redshifts higher than about 0.5.
The project how to do that and the selected sample of GRG candidates are presented
in Section~2. The imaging and spectroscopic observations conducting either with the SALT
or with the 1.9m SAAO telescope are described in Sections~3 and 4. Preliminary physical
and geometrical parameters of the sample sources are determined in Section~5, while
the equipartition energy density in their lobes (proportional to an internal lobe pressure)
vs. the sources' redshift and morphology of the GRGs vs. properties of the environment
are analyzed in Section~6.

\vspace{6mm}
\noindent
\underline{2. High-redshift GRGs and description of the project}

Since majority of known GRGs are at rather low redshifts (z$\leq$0.25), it was long
presumed that {\sl giants} did not exist at significantly higher redshifts, especially
because both the median and maximum source size were suggested to decrease as fast as
$(1+z)^{-3}$ (e.g. Gopal-Krishna \& Wiita 1987). Such a result could be explained by
systematic $(1+z)^{3}$ increase of density of an uniform IGM confining the source
size. Contrary, high-redshift GRGs might be tracers of relatively low density regions,
not fully virialized yet, existing at early cosmological epochs. Unfortunately,
unbiased samples of GRGs are very difficult to assemble because of surface-brightness
selection effects. At high redshifts, the diffused bridges of emission are likely to
be significantly influenced by the inverse-Compton losses against the cosmic background,
which would affect the appearance and identification of a GRG at these redshifts.

The first attempt to discover GRGs at $(z\geq 0.3)$ was made by Cotter et al.
(1996) who used the 7C radio survey to select Giant candidates and found 13
large double radio sources at redshifts up to $z\approx$0.9, 4 of them with
$D>$1 Mpc. A further few distant GRGs were found by Schoenmakers et al. (1998)
proving that such sources do exist at least out to $z\sim$1. In Table 1 (below)
we assemble all known {\sc Giants} with $z\geq$0.5 and $D\gtrsim$1 Mpc (calculated
with the cosmological parameters: $H_{0}$=71 km\,s$^{-1}$Mpc$^{-1}$, $\Omega_{m}$=0.27,
and $\Omega_{\Lambda}$=0.73.
This is worth emphasizing that only two sources in Table~1 have negative declinations.

\begin{table}[h]
\caption{Known {\sl giants} with z$\geq$0.5 and D$\gtrsim$1 Mpc}
\begin{tabular}{@{}lllcccrc}
\hline
IAU name & Survey    & z     & $\Theta$   & R & log\,P$_{1.4}$  & D & Id.\\
         &           &       & ["]        &[mag] & [W\,Hz$^{-1}$] & [kpc]\\
\hline
B0437-244  & PKS & 0.84  & 127 &      & 27.17 &  960 & QSO\\
B0654+482  & 7C  & 0.776 & 135 & 21.9 & 26.39 & 1002\\
J0750+656  &     & 0.747 & 220 &      & 26.40 & 1606 & QSO\\
B0821+695  & 8C  & 0.538 & 408 & 20.5 & 26.25 & 2576\\
B0854+399  & B2  & 0.528 & 166 &      & 26.71 & 1038\\
B1058+368  & 7C  & 0.750 & 150 & 21.3 & 26.53 & 1100\\
B1127-130  & PKS & 0.634 & 290 & 16.0 & 27.26 & 2033 & QSO\\
B1349+647  &3C292& 0.71  & 134 & 21.1 & 27.62 &  960\\
B1429+160  & MC3 & 1.005 & 167 &      & 26.88 & 1346 & QSO\\
B1602+376  & 7C  & 0.814 & 182 & $>$20& 26.59 & 1376\\
B1636+418  & 7C  & 0.867 & 130 & $>$20& 26.49 & 1004\\
B1834+620  & WNB & 0.519 & 220 &      & 26.89 & 1364\\
J1951+706  &     & 0.550 & 204 &      & 26.07 & 1300\\
\hline
\end{tabular}
\end{table}

In this project we attempt to enlarge significantly the above sample with
high-redshift GRGs on the southern hemisphere, and provide data sufficient
to enlighten the problem of the IGM pressure evolution with redshift. For this
purpose we selected a large sample of GRG candidates using the existing 1.4 GHz
radio surveys: FIRST (Becker et al. 1995) and NVSS (Condon et al. 1998). These
surveys, which provide radio maps of the sky made with two different angular
resolutions (5" and 45", respectively) at the same observing frequency, allow
(i) an effective removal of confusing sources, (ii) a reliable determination of
the sources' morphological type, and in many cases (iii) a determination of the
compact core component necessary for the proper identification of the source with
its host optical object.

In order to use a possibility of all-season optical observations with the SALT,
the entire sample consists of two parts (subsamples) located on the northern
and southern Galactic hemispheres:

\noindent
(NGH) lying between $\rm08^{h}20^{m}<RA(J2000)<16^{h}00^{m}$;\\ -14$\rm^{o}<$Dec(J2000)$<+13\rm^{o}$, and\\
(SGH) lying between $\rm21^{h}00^{m}<RA(J2000)<03^{h}15^{m}$;\\ -14$\rm^{o}<$Dec(J2000)$<+02\rm^{o}$.

Thus, an advance of the project is a uniform distribution of the targets along
hour angle enabling consecutive collection of their images and spectra. A crude
estimate of the source's redshift can be made from the Hubble diagram for GRGs
published by Schoenmakers et al. (1998). At $z$=0.5 it gives $R\approx$20 mag
(the apparent magnitudes in Table~1 confirm this). On the other hand, a source
with the linear projected size of 1 Mpc at $z$=0.5 will have an angular size
of about 165". Therefore the GRG candidates were chosen to fulfill the following
selection criteria: they have

\noindent
(1) the morphological type of FRII,\\
(2) an angular separation between the brightest regions on the maps $\Theta>165$",\\
(3) 1.4-GHz flux density on the NVSS map $S>$50 mJy in the NGH subsample,
and $S>$40 mJy in the SGH subsample, and\\
(4) is optically identified with an object fainter than 20 mag in the $R$-band.

The above criteria selected rather small number of high-redshift GRG candidates,
especially in the NGH subsample. Therefore in both subsamples we include also
other radio source candidates with host galaxies brighter than 20($R$) mag which
will be analyzed as comparison closer GRGs. The optical spectroscopy of these latter
sources can be conducted with smaller telescopes. In Table 2 we list numbers of the
GRG candidates in both subsamples. The observational data on
34 of 39 radio sources in the subsample NGH and 12 of 41 in the subsample SGH, i.e.
those for which preliminary physical parameters could be determined using the data
collected during the P-V phase, are given in Table~3.
The consecutive columns of Table 3 give:

\noindent
Col. 1: IAU name at epoch J2000;\\
Col. 2: Fanaroff-Riley morphological type. ``D'' indicates a diffuse morphology of the radio
lobes without any detectable hot spot(s);\\
Col. 3: 1.4 GHz total flux density in mJy from the NVSS survey;\\
Col. 4: radio spectral index between 1.4 and 4.9 GHz;\\
Col. 5: 1.4 GHz flux density of the radio core from the FIRST survey;\\
Col. 6: largest angular size in arcsec;\\
Col. 7: optical identification. ``G'' - galaxy, ``Q'' - quasar;\\
Col. 8: its apparent R-band magnitude from the DSS data base;\\
Col. 9: redshift.  A value in parenthesis is the photometric redshift estimate calculated
from the log\,$z$--$R$ relation (Hubble diagram) derived from the spectroscopic
redshift and K-corrected  magnitudes of the galaxies in Table 3. Our relation
is consistent with the Hubble diagram for a majority of known GRGs published by Lara et al.
(2001) [their equation (1) and figure 4];\\
Col. 10: redshift references. (1) our SAAO observations, (2) our SALT observations,
(3) Sloan Digitized Sky Survey [SDSS] Adelman-McCarthy et al. (2007), (4) Jones et al.
(2005), (5) Colless et al. (2001), (6) Machalski et al. (2001), (7) Bhatnagar et al. (1998),
(8) Stickel et al. (1994), (9) Becker et al. (2001). 

\begin{table}[h]
\caption{Content of the subsamples}
\begin{tabular}{lllll}
\hline
Subsample & Redshift already & to be determined & to be determined & Total\\
          & available        & with the SALT  & with other tel.\\
\hline
NGH  & 13 & 16 & 10 & 39\\
SGH  & \, 7 & 29 & \, 5 & 41\\
\hline
\end{tabular}
\end{table}

\begin{table}[]
\footnotesize{
\caption{The sample. The entries determined in this paper are marked in bold face.}
\begin{tabular}{llrrrrlllr}
\hline
Name     & FR   & S$_{1.4}$ & $\alpha$ & S$_{core}$ & LAS & Opt. & R$_{\rm DSS}$ & $z$ &
Ref.\\
         & type & [mJy]   & & [mJy] & ["]  & id.  & [mag] & & to $z$\\
\hline
{\bf NGH:}\\
J0824+0140   & IID &   52 &     &  5.9 &  290 & G & 16.91 &{\bf 0.2125} & 1\\
J0903+1208   & II  &  113 &     &$<$1. &  320 & G & 18.73 & (0.27) \\ 
J0922+0919   & IID &  104 &     &  1.3 &  434 & G & 18.11 & (0.23) \\
J0925$-$0114 & IID &   71 &     & 12.8 &  832 & G & 12.30 & 0.0732 & 4\\
J0947$-$1338 & II  &  977 & 0.85& 20.7 & 1514 & G & 12.70 &{\bf 0.0800} & 1\\
J1005$-$1315 & IID &   58 &     &      &  260 & G & 19.34 & (0.31) \\
J1014$-$0146 & II  &  225 & 0.76& 20.4 &  292 & G & 16.54 & 0.1986 & 5\\
J1018$-$1240 & I/II&  276 & 1.27& 24.8 &  570 & G & 14.66 & 0.0777 & 6\\
J1021$-$0236 & II  &  218 & 0.99&  7.6 &  357 & G & 18.50 &{\bf 0.2917} & 2 \\
J1021+1217   & II  &  138 &     & 16.9 &  865 & G & 15.49 & 0.1294 & 3\\
J1021+0519   & II  &  135 &     &  5.8 &  835 & G & 16.26 & 0.1562 & 3\\
J1048+1108   & IID &  235 & 1.02&  3.2 &  294 & G & 15.71 & 0.1570 & 3\\
J1049$-$1308 & IID &  193 & 0.89&  ?   &  420 & G & 16.05 & (0.15) & \\
J1058+0812   & II  &   49 &     &$<$1. &  367 & G & 19.74 & (0.33) & \\
J1101$-$1053 & II  &   92 &     &$<$6. &  852 & G & 16.19 & (0.15) & \\
J1108+0202   & II  &  209 &     &  ?   &  556 & G & 15.57 & 0.1574 & 3\\
J1126$-$0042 & II  &  166 &     &  1.7 &  239 & G & 18.10 &{\bf 0.3317} & 2\\
J1130$-$1320 & II  & 1136 & 1.03& 18.2 &  297 & Q & 15.98 & 0.6337 & 7\\
J1213$-$0500 & IID &$>$38 &     & 14.3 &  535 & G & 15.69 &{\bf 0.0857} & 1\\
J1253$-$0139 & II  &  171 & 0.61&$<$1. &  231 & G & 18.33 & (0.24) & \\
J1328$-$0129 & II  &  343 &     & 16.6 &  323 & G & 16.91 & 0.1513 & 3\\
J1328$-$0307 & II  &  201 & 1.2 &  8.8 &  810 & G & 16.82 & 0.0860 & 3,6\\
J1334$-$1009 & II  & 1893 & 0.67& 100. &  850 & G & 13.58 & 0.0838 & 4,8\\
J1354$-$0705 & II  &   95 &     &  2.2 &  240 & G & 17.30 & (0.19) & \\
J1411+0619   & II  &  115 & 0.77& 22.0 &  310 & G & 20.00 &{\bf 0.3591} & 2\\
J1420$-$0545 & II  &   89 &     &  2.7 & 1045 & G & 19.65 & (0.32) \\
J1445+0932   & II  &  144 & 0.95&  6.4 &  315 & G & 14.89 &{\bf 0.094} & 1\\
J1445$-$0540 & II  &  235 &     & 12.6 &  420 & G & 19.44 &{\bf 0.3666} & 2\\
J1457$-$0613 & II  &  507 &     &      &  253 & G & 16.64 & 0.1671 & 6\\
J1459$-$0432 & IID &   62 &     & 0.73 &  405 & G & 18.76 & (0.27) & \\ 
J1520$-$0546 & II  &   63 &     & 10.7 & 1320 & G & 12.35 &{\bf 0.0607} & 1\\
J1528+0544   & II  &  247 & 1.05& 20.3 &  824 & G & 11.5  &{\bf 0.040} & 1\\
J1540$-$0127 & II  &  209 & 1.06& 10.3 &  295 & G & 16.57 & 0.1490 & 3\\
J1543$-$0112 & II  &  117 & 0.81&  0.5 &  175 & G & 20.78 &{\bf 0.3680} & 2\\
{\bf SGH:}\\
J2234$-$0224 & II  &   73 &     &  6.1 &  198 & Q & 18.63 & 0.55   & 9\\
J2239$-$0133 & II  &  134 &     &  1.8 & 1225 & G & 13.45 &{\bf 0.0865} & 1\\
J2345$-$0449 & IID &  162 &     &  3.5 & 1025 & G & 13.4  & 0.0757 & 4\\
J0010$-$1108 & IID &   55 &     &  5.4 &  550 & G & 12.57 & 0.0773 & 3\\
J0037+0027   & II  &  134 & 1.08&  2.1 &  299 & G &{\bf 20.55} & (0.42) \\
J0039$-$1300 & II  &  197 & 0.70&      &  640 & G & 12.93 &{\bf 0.1060} & 1\\
J0042$-$0613 & II  & 1204 & 0.91&      &  390 & G & 16.47 &{\bf 0.123} & 1\\
J0129$-$0758 & I/II&  102 &     &      &  720 & G & 14.0  & 0.0991 & 4\\
J0134$-$0107 & I/II&  287 &  ?  &      &  820 & G & 12.25 & 0.0790 & 3\\   
J0202$-$0939 & II  &  431 & 0.87&  0.4 &  154 & G &{\bf 20.2} & (0.37) \\
J0300$-$0728 & II  &   51 &     &  1.2 &  300 & G & 20.44 & 0.4905 & 3\\
J0313$-$0632 & II  &  165 & 1.0 &      &  195 & Q & 19.20 & 0.389  & 3\\     
\hline
\end{tabular}}
\end{table}

\newpage
\noindent
\underline{3. Optical imaging and identification}

\vspace{2mm}
\noindent
{\sl 3.1. DSS identification}

To identify the GRG candidates whose radio core is known with a host optical object,
first at all we used the DSS, i.e. the digitized POSS and
UKST surveys. For all NGH sources with cores, we have identified their cores with
galaxies; only one sample source is known radio quasar (J1130$-$1320). For the remaining
NGH candidates -- the galaxies given in Table~3, being the only objects  in a middle of
the radio structure, thus are very probable identifications. Quite different situation
is in the SGH subsample whose sky area is almost not covered by the FIRST survey.
Therefore for a certain optical identifications high-resolution radio observations
are necessary. The SGH sources, where at the position of detected radio core no optical
object had been visible in the DSS data base, were scheduled for a deep detection with
the SALT.   

\vspace{2mm}
\noindent
{\sl 3.2. Preliminary deep imaging  and photometry with the SALT}

\vspace{2mm}
During the {\sl Performance Verification} (P-V)  phase two blank fields in the SGH
subsample were observed using the SALTICAM. Series of 60sec-long exposures were taken
through the $B$, $V$ and $R$ filters. The frames were corrected for overscan, 
cross-talk and mosaiced. Because flat-field frames were not available for the tracker
position close to the objects' position, the surface functions were fitted and then
subtracted from the frames in order to obtain possibly flat and uniform background.
Instrumental magnitudes of objects in the frames were determined using the aperture
method.
Apparent magnitudes of these objects were estimated using magnitudes of stars in each
field for which their standard magnitudes were available from the APM catalogue
(http://www.ast.cam.ac.uk/$\sim$apmcat/). The astrometric calibration was done transforming
the instrumental coordinates of these stars in a given frame into their sky coordinates
in the DSS data base. These coordinates were not corrected for the proper motion and
parallax.

The deep optical images of the field around the sources J0037+0027 and J0202$-$0939
are shown in Figs.\,1a, b. Overlaid are intensity-contours of the radio maps made with
the combined FITS data of the FIRST and NVSS surveys. On both images a faint galaxy
coincides with the radio core. The resulting magnitudes and their errors are given in
Table~4.

\vspace{2mm}
\noindent
{\sl 3.3. Auxiliary photometry of J1420$-$0545}

Because of a suspicion that the radio galaxy J1420$-$0545 can be the largest GRG
observed up to now (cf. Introduction), we made an auxiliary optical photometry for
this galaxy. The observations were made using the 60cm telescope of the Pedagogical
Academy at Mt. Suhora.
The resulting magnitudes in the $UBVRI$ photometric system, given in Table 4, are
corrected for the atmospheric extinction only. As the galactic latitude of this
galaxy is 50 degr, the foreground Galactic extinction is close to zero (Sandage 1972).
In order to be consistent with the DSS data, these magnitudes are not K-corrected.
Our $R$-band magnitude is thus identical as the corresponding value in Table 3 taken
from the DSS data base. The optical spectral index of about 4.0 resulting from our
VRI magnitudes also suggests a redshift higher than 0.3 for any of the evolutionary
models for early-type galaxies (Bruzual 1983).

\begin{figure}[h]
\begin{center}
\includegraphics{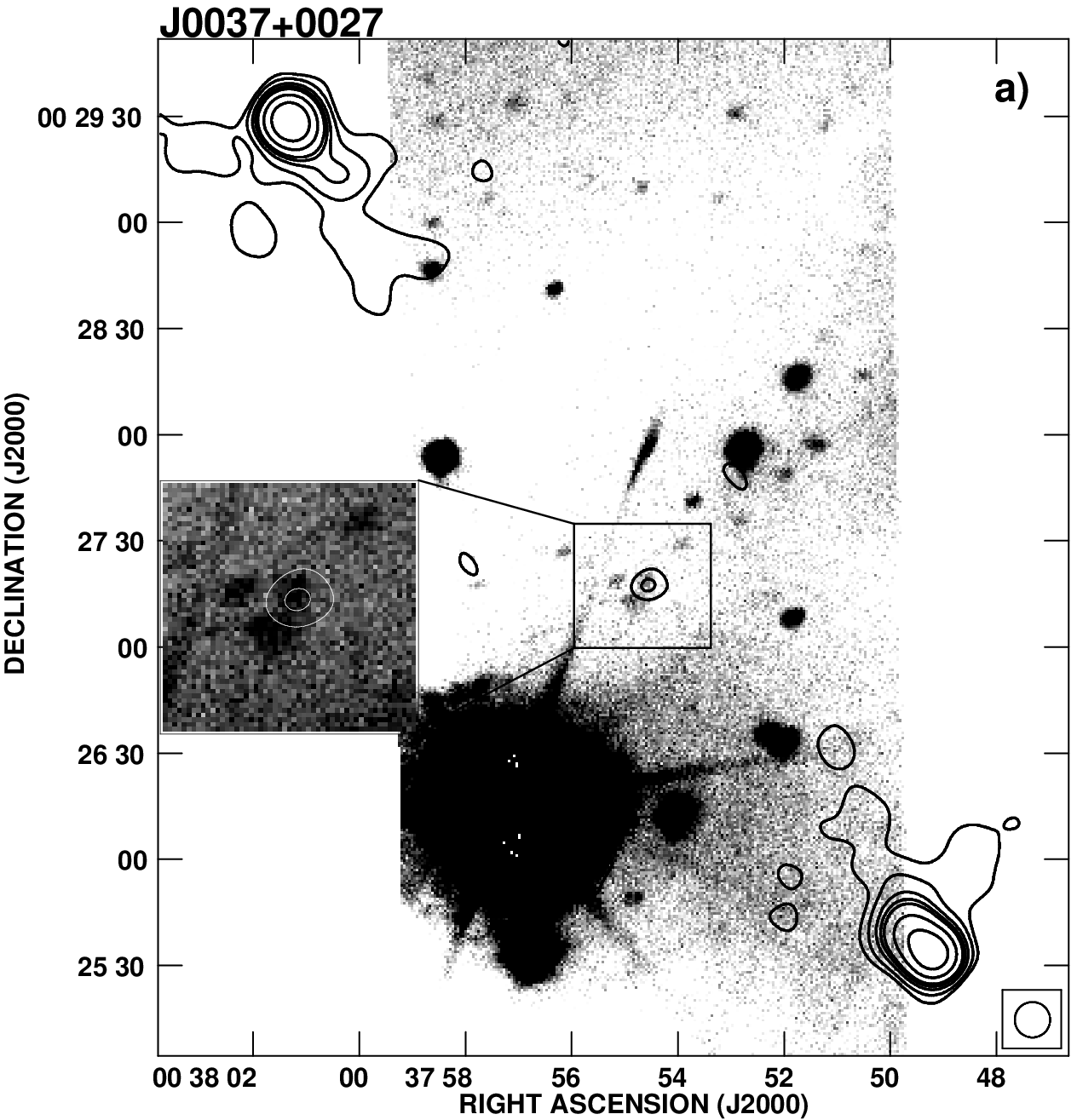}
\includegraphics{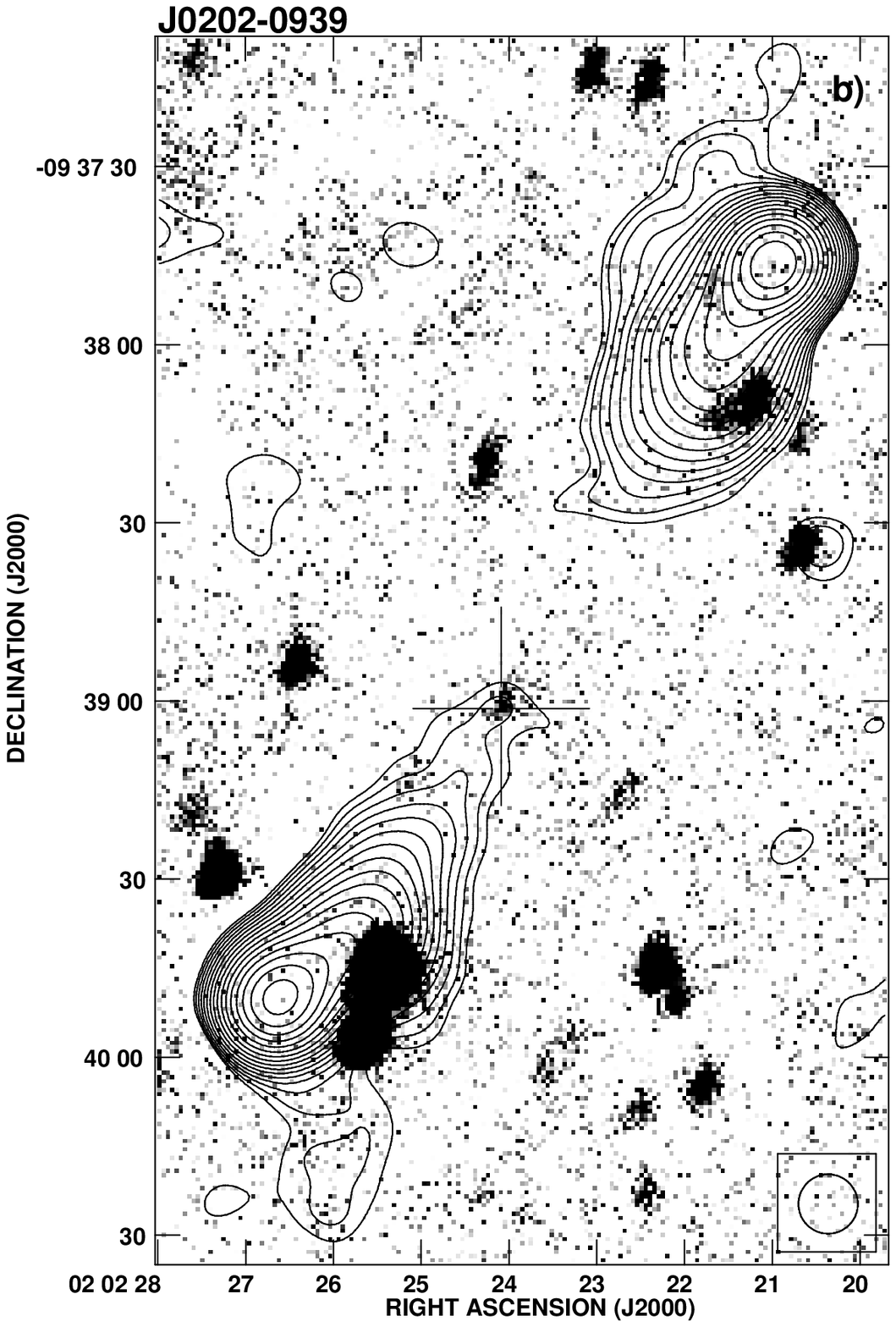}
\vspace{110mm}
\caption{Deep images of the optical fields taken with the SALTICAM: {\bf a)} around
the source J0039+0027 and {\bf b)} around the source J0202$-$0939. The solid contours
indicate the 1.4 GHz brightness distribution observed in the NVSS + FIRST surveys}
\end{center}
\end{figure}

\begin{table}[h]
\caption{$IRVB$ magnitudes of the detected galaxies}
\begin{tabular}{lcccc}
\hline
Source  & $I$[mag] & $R$[mag] & $V$[mag] & $B$[mag]\\
\hline
J0037+0027  &-- & 20.55$\pm$0.16  & 21.69$\pm$0.15  & 23.12$\pm$0.13 \\
J0202$-$0939  &-- &   --          & 21.36$\pm$0.15  & 22.89$\pm$0.12\\
J1420$-$0545  & 18.44$\pm$0.08 & 19.65$\pm$0.08 & (21.0$\pm$0.2) & --\\
\hline
\end{tabular}
\end{table}
 
\noindent
\underline{4. Optical spectroscopy}

\vspace{3mm}
\noindent
{\sl 4.1. Low-resolution spectroscopy with the SALT}

\vspace{2mm}
In order to determine physical parameters of a given sample source, e.g. its radio
and optical luminosity, linear size and volume, energy density, etc., a distance
to the host galaxy (or QSO) must be known which can be calculated from its redshift.
Therefore, using the low-resolution spectrograph (PFIS) during the P-V observations
we attempted to detect optical spectrum and determine redshifts of 6 faint host galaxies.
The observations were made in series between May 30th 2006 and June 24th 2006. The
spectra were obtained with the PFIS in Grating Spectroscopy mode using the standard
surface-relief grating SR300, slit width of 1.5 arcsec, granting angle of 7 deg, and
the binning of 2$\times$2. These settings resulted in the typical useful spectral range
about 3800\AA --8500\AA \hspace{1mm} and the dispersion of 3.05\AA \hspace{1mm}per pixel.
Because of the certain space between the CCD chips there is a gap between 
6280\AA \hspace{1mm}and 6440\AA \hspace{1mm} in all the SALT spectra. The pre-reduced
(i.e. bias and overscan subtracted,
trimmed, cross-talk corrected, mosaiced and cosmic-ray removed) files were further
reduced using the standard IRAF tasks. All the spectra were wavelength calibrated
using the CuAr arc lamp exposures and one-dimensional spectra were extracted. Gaussian
profiles were fitted to the lines visible in the spectra and the wavelengths
corresponding to the profiles' centers were used to calculate the redshift. 

\vspace{3mm}
\noindent
{\sl 4.2. Low-resolution spectroscopy with the 1.9m SAAO telescope}

\vspace{2mm}    
9 host galaxies brighter than about 16\,mag ($R$) were observed with the 1.9m telescope
equipped with the Cassegrain Grating Spectrograph in the long-slit mode. The galaxies
were observed with low-resolution grating 8, slit width of 300$\mu$m (2.5 arcsec)
and the grating angle of 14.7\,deg resulting in the spectral range 5300\AA --7600\AA \hspace{1mm}
and the dispersion of 1.66\AA \hspace{1mm}  per pixel. Exposures of the CuAr arc lamp were made every
30 min. Every night selected spectrophotometric standards were also observed.

Reduction of the spectra were carried out using the standard IRAF longslit package. 
All the spectra were corrected for bias, flat-field and cosmic-rays, as well as wavelength
and flux-density calibrated. In order to obtain the resulting spectra with a better S/N
ratio, all calibrated one-dimensional spectra of a given galaxy were combined.

\begin{table}[t]
\footnotesize{
\caption{Lines detected and redshift of the galaxies}
\begin{tabular}{llclllcl}
\hline
Source & Line/band & $\lambda_{obs}$ & $z$ & Source & Line/band & $\lambda_{obs}$ & $z$\\
       & detected         &[\AA]  & & &  detected        &[\AA]\\
\hline
J0039$-$1300 & Mg band    & 5725.9 & 0.1060      & J1411+0619 & [OII]3727  & 5065.9 & 0.3592\\
             & NaD band   & 6519.9 & 0.1058      &            & CaII\,3935 & 5350.7 & 0.3598\\
             & [NII]6550  & 7244.0 & 0.1060      &            & CaII\,3970 & 5396.0 & 0.3592\\
             & [NII]6585  & 7284.0 & 0.1061      &            & G band     & 5853.8 & 0.3598\\
             & [SII]6718  & 7432.0 & 0.1062      &            & Mg band    & 7032.6 & 0.3584\\
             & [SII]6733  & 7446.0 & 0.1059      &            & NaD band   & 8008.2 & 0.3582\\
             &   &   &$\overline{0.1060}\pm$0.0002 &  & & & $\overline{0.3591}\pm$0.0008\\
\\
J0042$-$0613 & Mg band    & 5810.5 & 0.1224      & J1445+0932 &  Mg band   & 5661.5 & 0.0936\\     
             & NaD band   & 6623.5 & 0.1234      &            & NaD band   & 6450.8 & 0.0941\\
             &   &   &$\overline{0.123}\pm$0.0010  & & & &  $\overline{0.094}\pm$0.0010\\
\\
J0824+0140   & H$\beta$   & 5894.2 & 0.2125      & J1445$-$0540 & [OII]3727 & 5094.1 & 0.3668\\
             & [OIII]4959 & 6014.0 & 0.2125      &              & [OIII]4959& 6777.2 & 0.3664\\
             & [OIII]5007 & 6070.6 & 0.2124      &              & [OIII]5007& 6842.3 & 0.3665\\
             &  &  & $\overline{0.2125}\pm$0.0001 & & & &  $\overline{0.3666}\pm$0.0003\\
\\
J0947$-$1338 & Mg band    & 5591.1 & 0.0800      & J1520$-$0546 & Mg band   & 5489.7 & 0.0604\\
             & NaD band   & 6367.0 & 0.0799      &              & NaD band  & 6252.5 & 0.0605\\
             & [NII]6550  & 7072.9 & 0.0798      &              & [NII]6585 & 6986.5 & 0.0610\\
             & H$\alpha$  & 7089.8 & 0.0803      &              & [SII]6733 & 7142.5 & 0.0608\\
             & [NII]6585  & 7112.2 & 0.0801      &  &  &  & $\overline{0.0607}\pm$0.0003\\
             & [SII]6718  & 7255.4 & 0.0800\\
             &  &  & $\overline{0.0800}\pm$0.0002\\
\\
J1021$-$0236 & [OII]3727  & 4814.9 & 0.2919      & J1528+0544   & Mg band   & 5381.4 & 0.0395\\
             & [OIII]5007 & 6466.5 & 0.2915      &              & NaD band  & 6135.4 & 0.0406\\
             &  &  & $\overline{0.2917}\pm$0.0005 &  &  &  &$\overline{0.040}\pm$0.0010\\
\\
J1126$-$0042 & [OII]3727  & 4965.1 & 0.3322      & J1543$-$0112 & [OII]3727  & 5100.4 & 0.3685\\
             & [OIII]4960 & 6602.0 & 0.3310      &              & [OIII]4364 & 5971.8 & 0.3684\\ 
             & [OIII]5007 & 6665.7 & 0.3313      &              & [OIII]4960 & 6784.4 & 0.3678\\
             & H$\alpha$  & 8742.9 & 0.3321      &              & [OIII]5007 & 6848.8 & 0.3678\\
             &  &  & $\overline{0.3317}\pm$0.0007 &  &  &  &$\overline{0.3681}\pm$0.0005\\
\\
J1213$-$0500 & Mg band    & 5618.7 & 0.0853      & J2239$-$0133 & Mg band    & 5622.9 & 0.0861\\
             & NaD band   & 6399.0 & 0.0853      &              & NaD band   & 6406.5 & 0.0866\\
             & H$\alpha$  & 7127.4 & 0.0860      &              & H$\alpha$  & 7157.3 & 0.0865\\
             & [NII]6585  & 7153.6 & 0.0863      &  &  &  &  $\overline{0.0865}\pm$0.0006\\
             &  &  & $\overline{0.0857}\pm$0.0005\\
\hline
\end{tabular}}
\end{table}

\vspace{3mm}
\noindent
{\sl 4.3. The resulting spectra}

\vspace{2mm}
The final, low-resolution, 1--D spectra of 14 observed galaxies are shown in Figs. 2a--d.
The emission lines and/or absorption bands detected, as well as the resultant redshift of
these features are listed in Table~5. Because the spectra taken with the SALT were not
flux-density calibrated,  relevant ordinates in Figs. 2 give the normalized counts only.
Most of the flux-density calibrated spectra are typical of elliptical galaxies whose
continuum emission with the broad absorption bands is dominated by evolved giant stars. 
The emission lines mostly detected are [OII]$\lambda$3727 and [OIII]$\lambda$4960 and
$\lambda$5007. Other forbidden lines present in a few spectra are [NII]$\lambda$6585,
as well as [NII]$\lambda$6718 and $\lambda$6733. Also weak Balmer lines are detected
in five of the spectra, however the H$\alpha$ line redshifted by more than 1.25 was
usually beyond the wavelength range observed.

\renewcommand{\thefigure}{2a}
\begin{figure}[t]
\begin{center}
\includegraphics{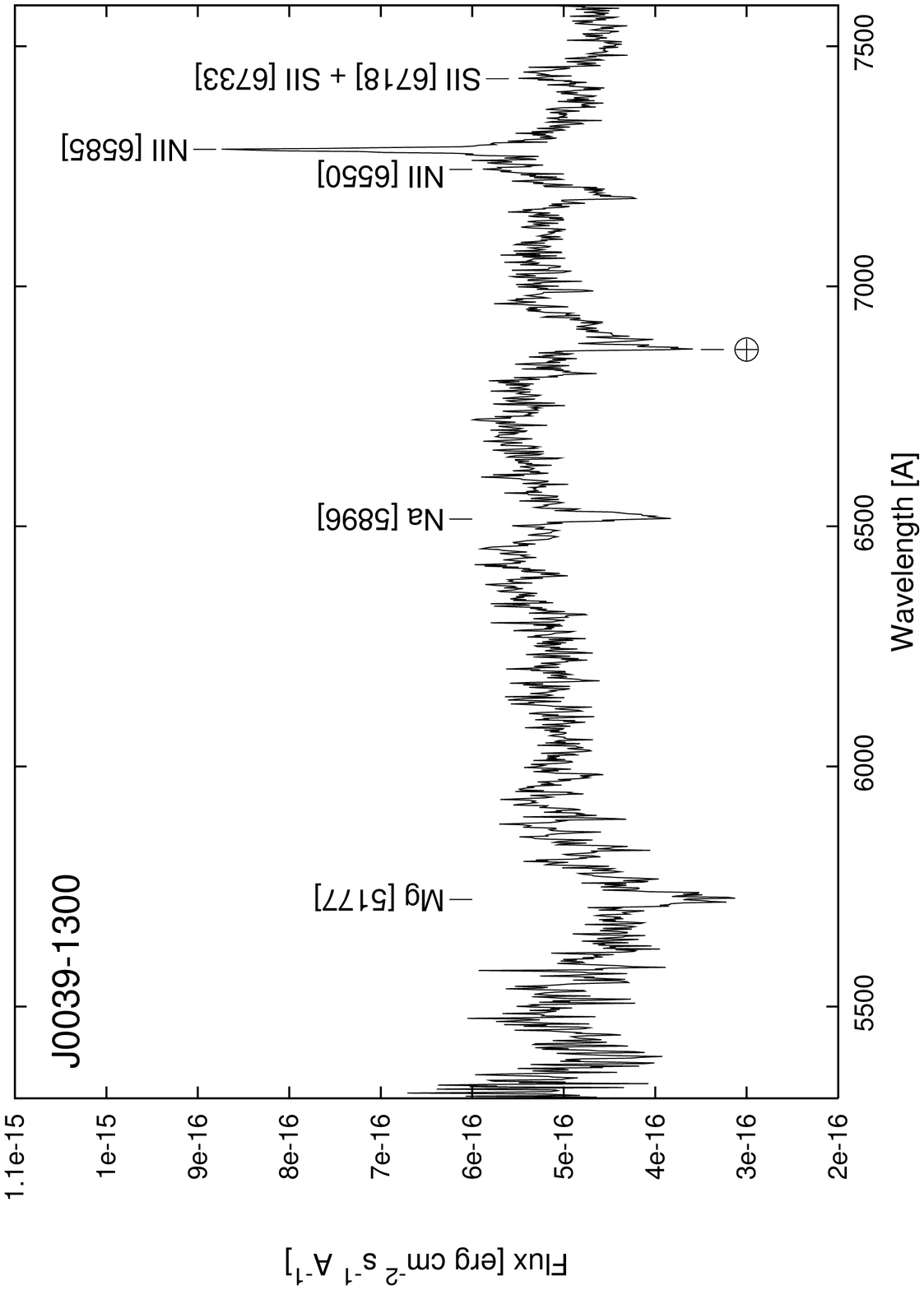}
\includegraphics{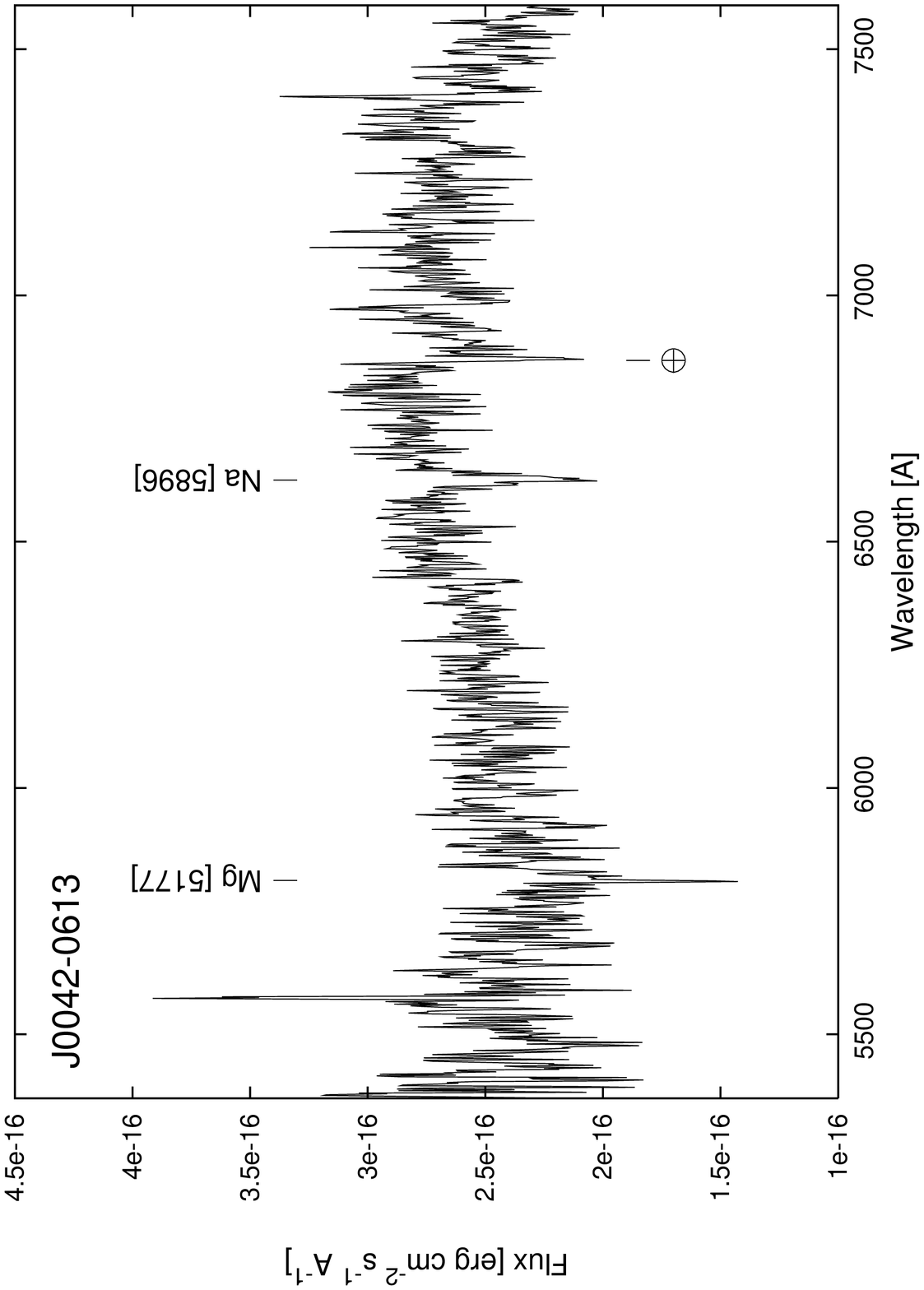}
\includegraphics{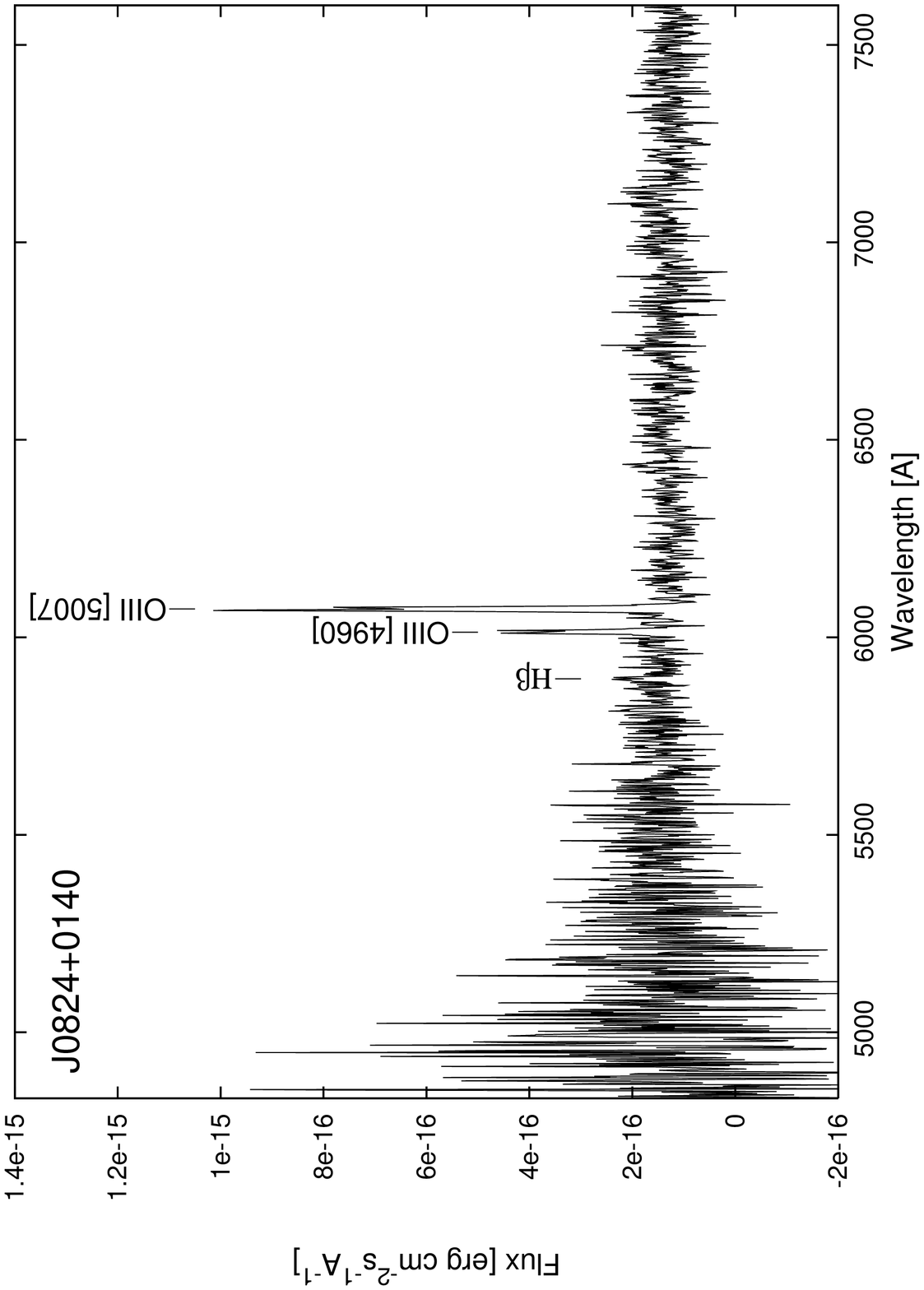}
\includegraphics{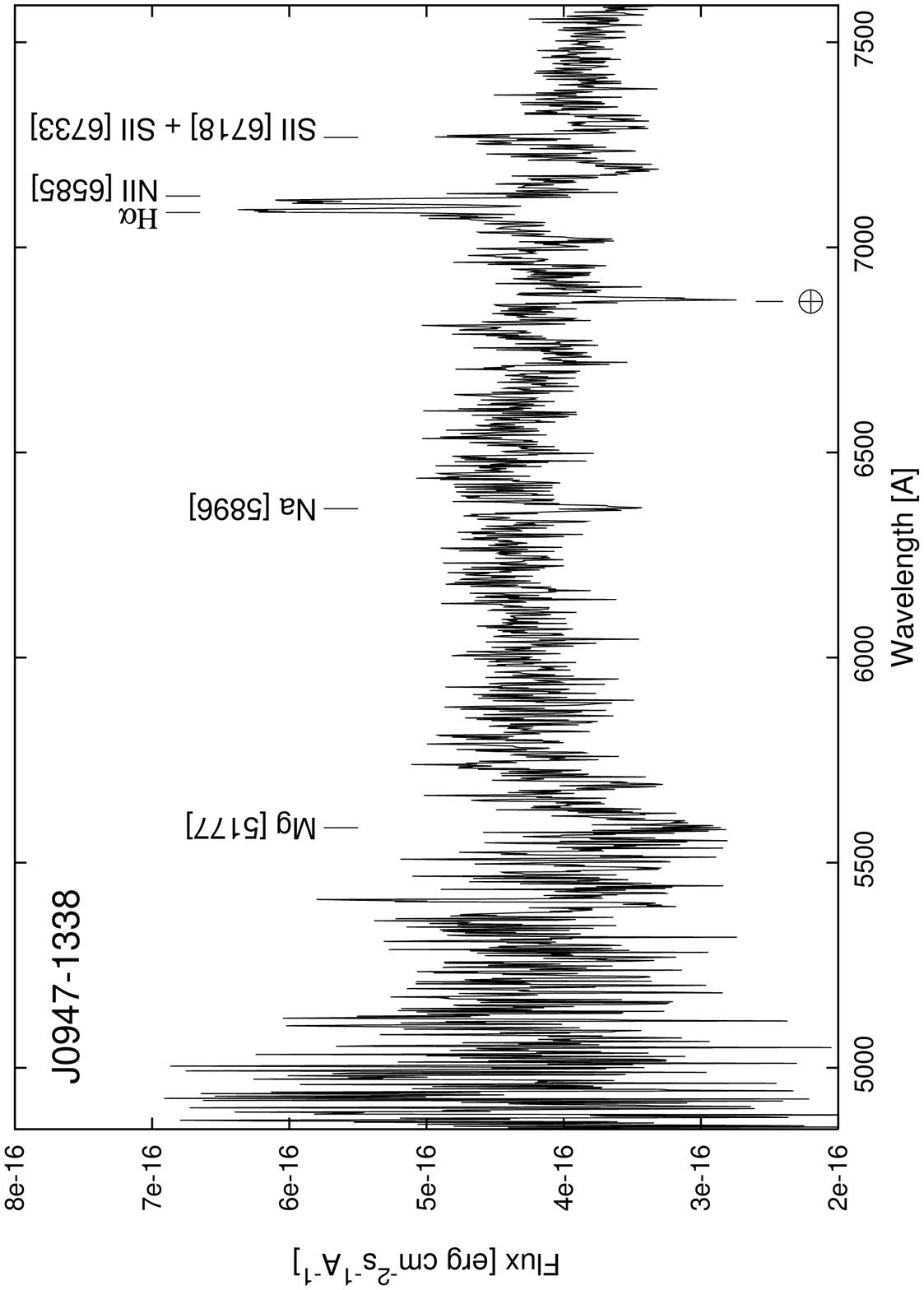}
\vspace{210mm}
\caption{Optical low-resolution spectra of the sources: J0039$-$1300, J0042$-$0613,
J0824$+$0140, J0947$-$1338, whose IAU names are marked in each panel}
\end{center}
\end{figure}

\renewcommand{\thefigure}{2b}
\begin{figure}[t]
\begin{center}
\includegraphics{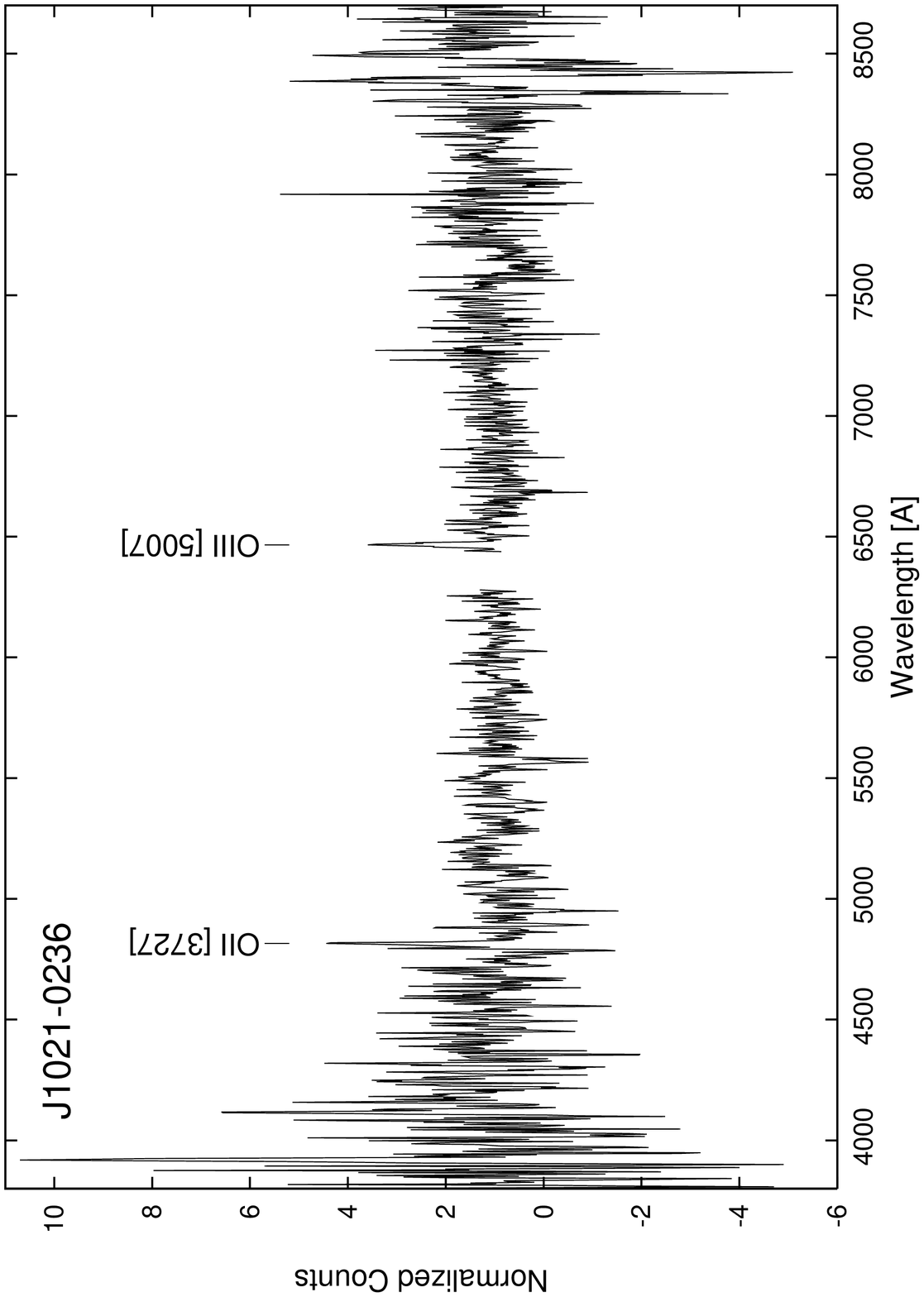}
\includegraphics{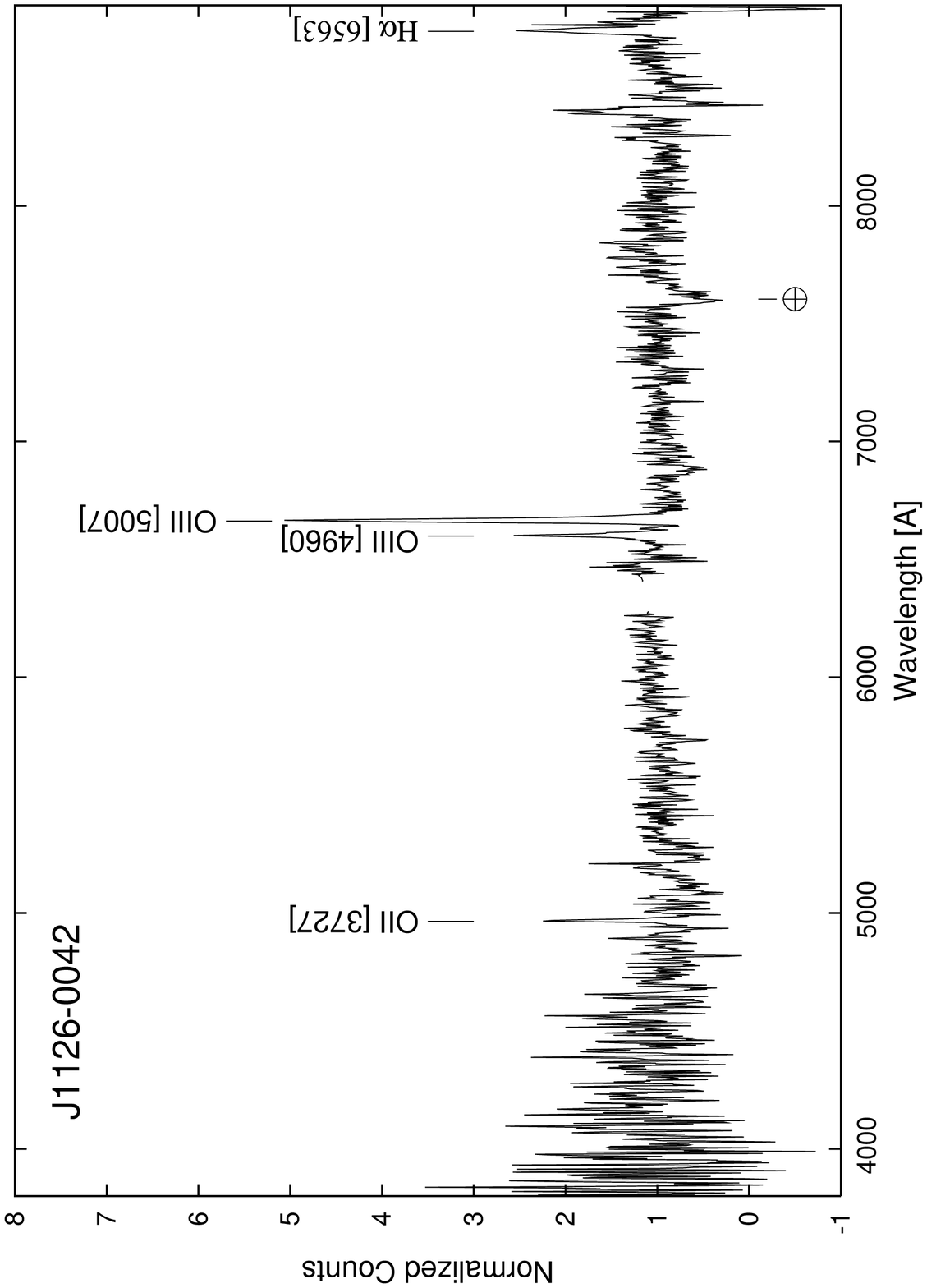}
\includegraphics{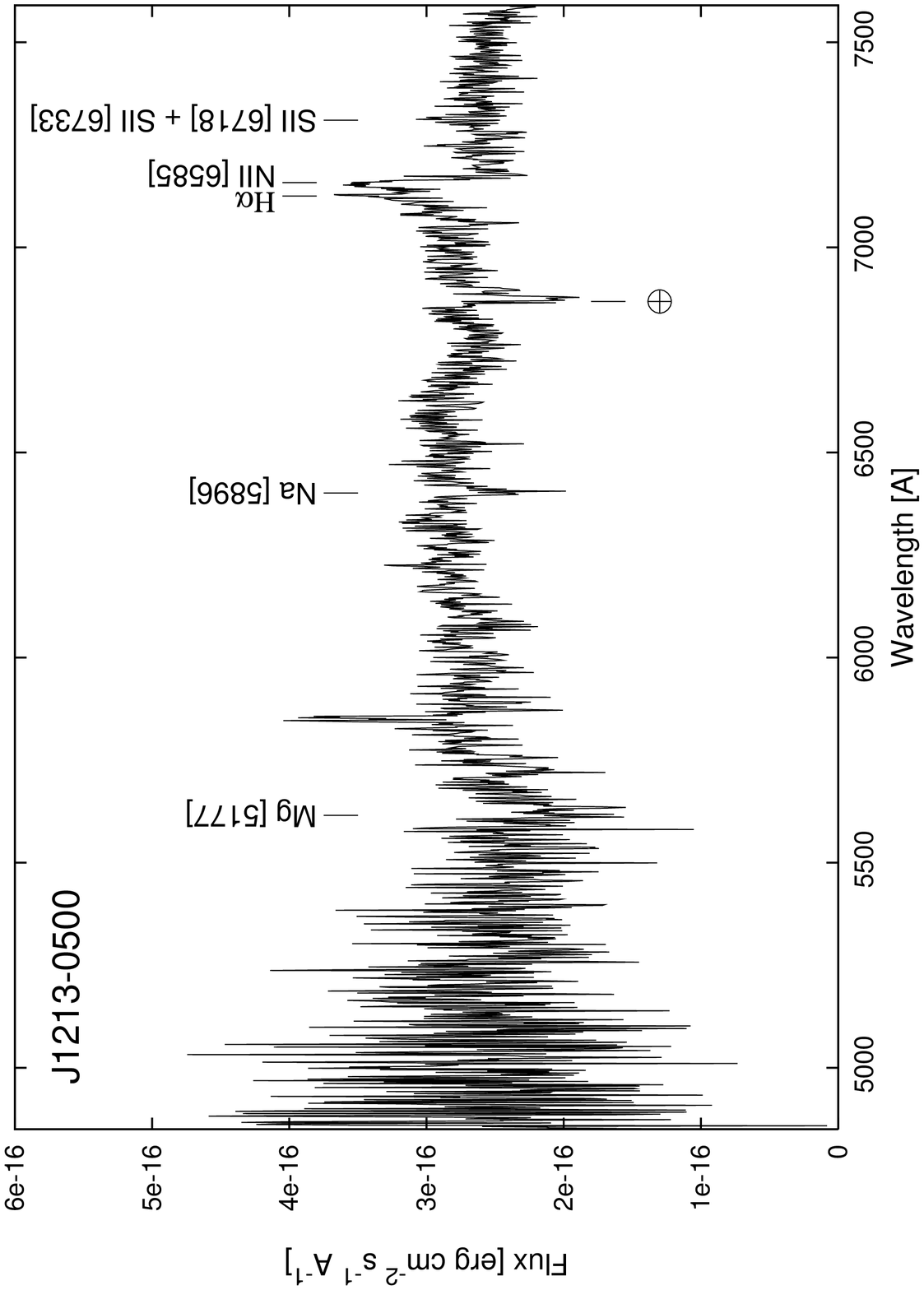}
\includegraphics{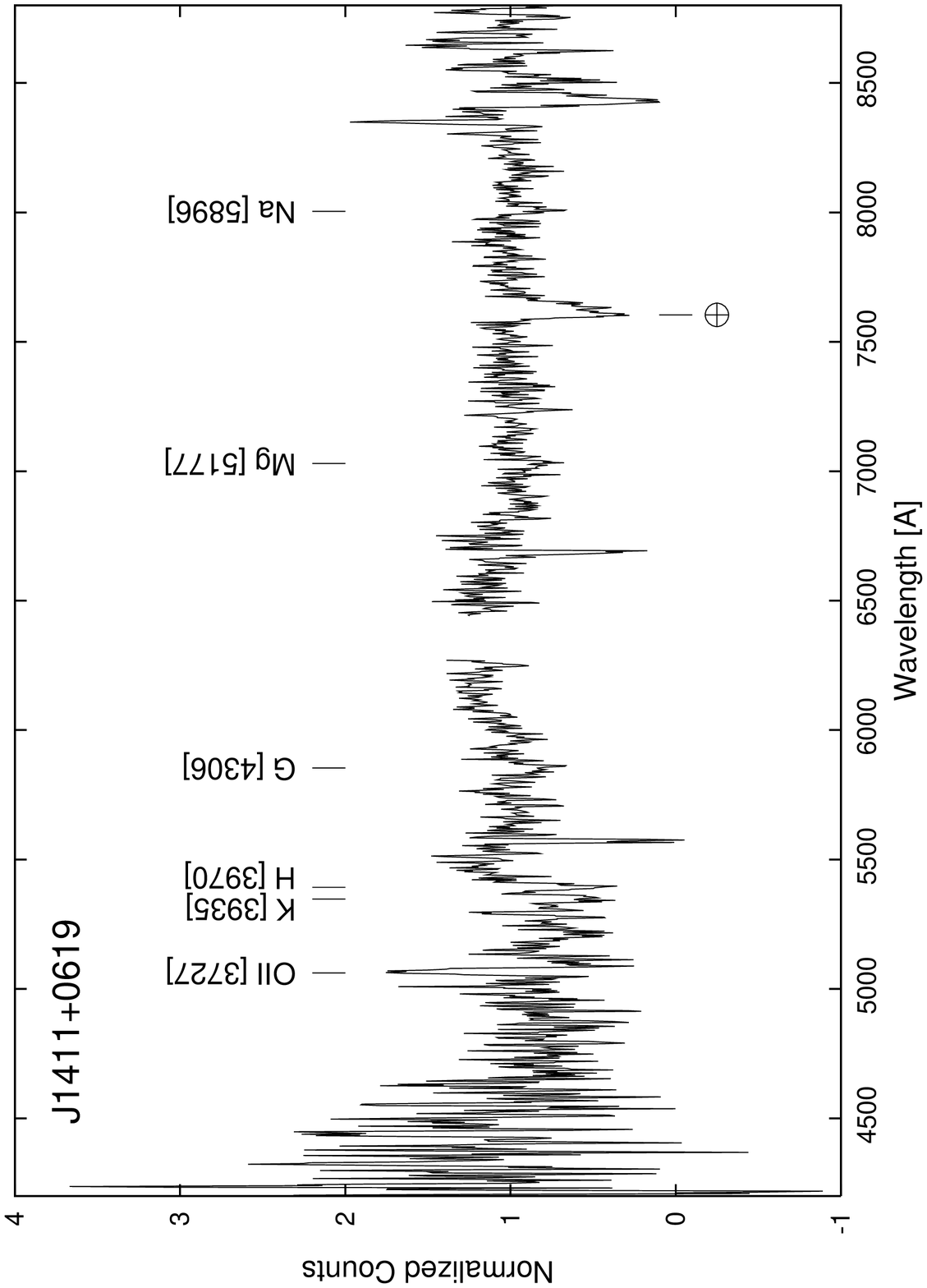}
\vspace{210mm}
\caption{The same as in Fig.~2a but for the sources: J1021$-$0236, J1126$-$0042, 
J1213$-$0500, J1411$+$0619}
\end{center}
\end{figure}

\renewcommand{\thefigure}{2c}
\begin{figure}[t]
\begin{center}
\includegraphics{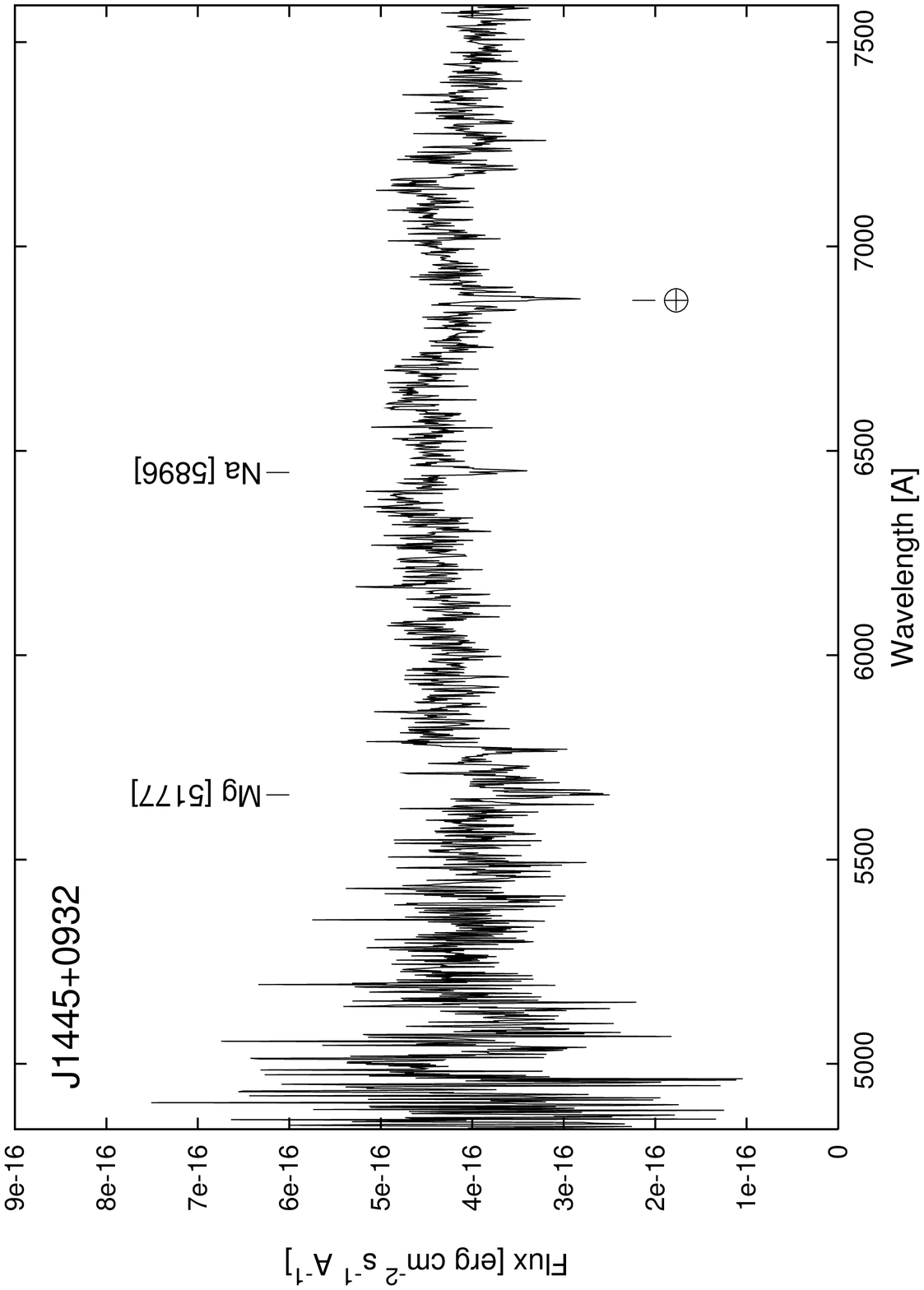}
\includegraphics{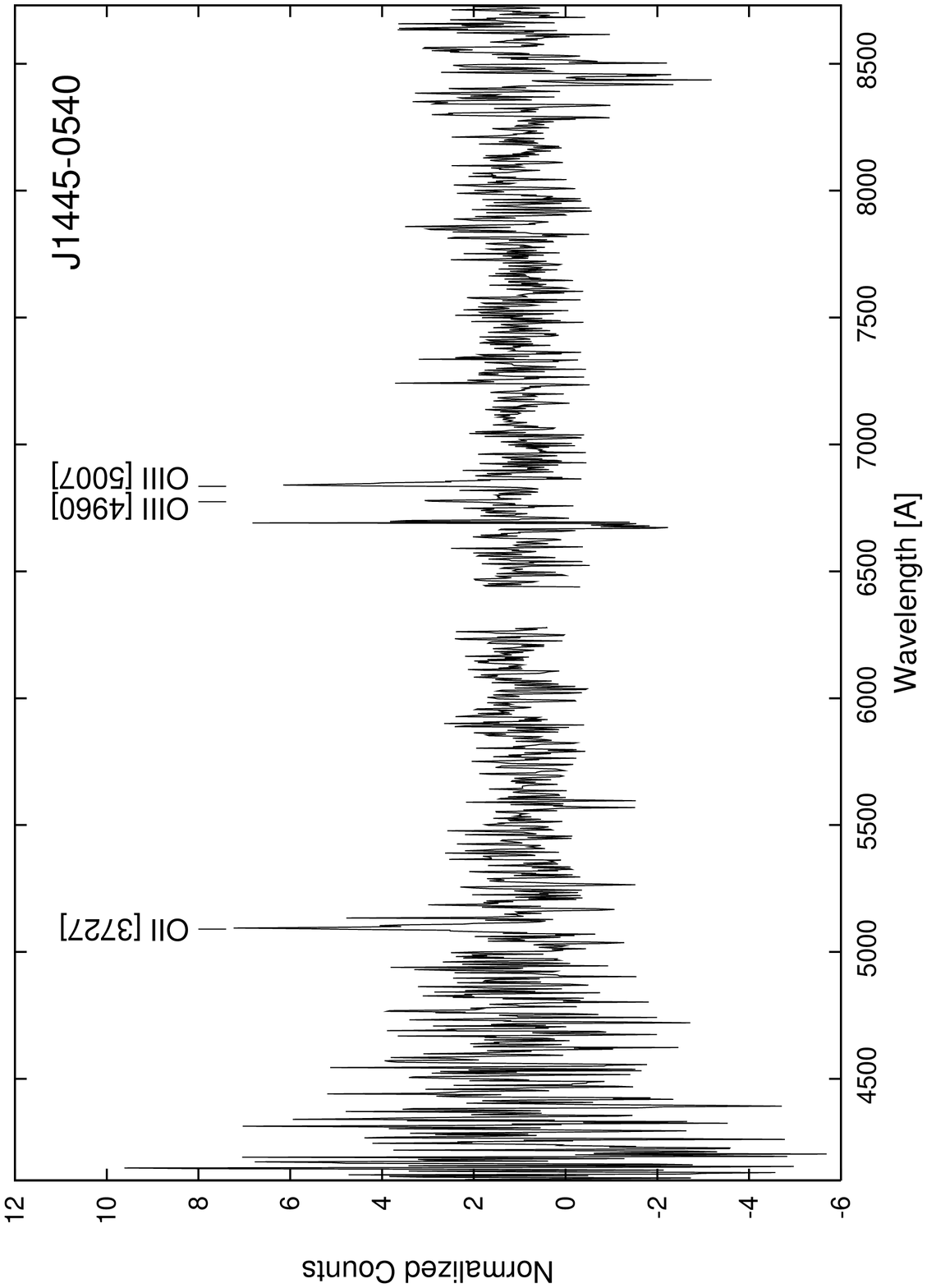}
\includegraphics{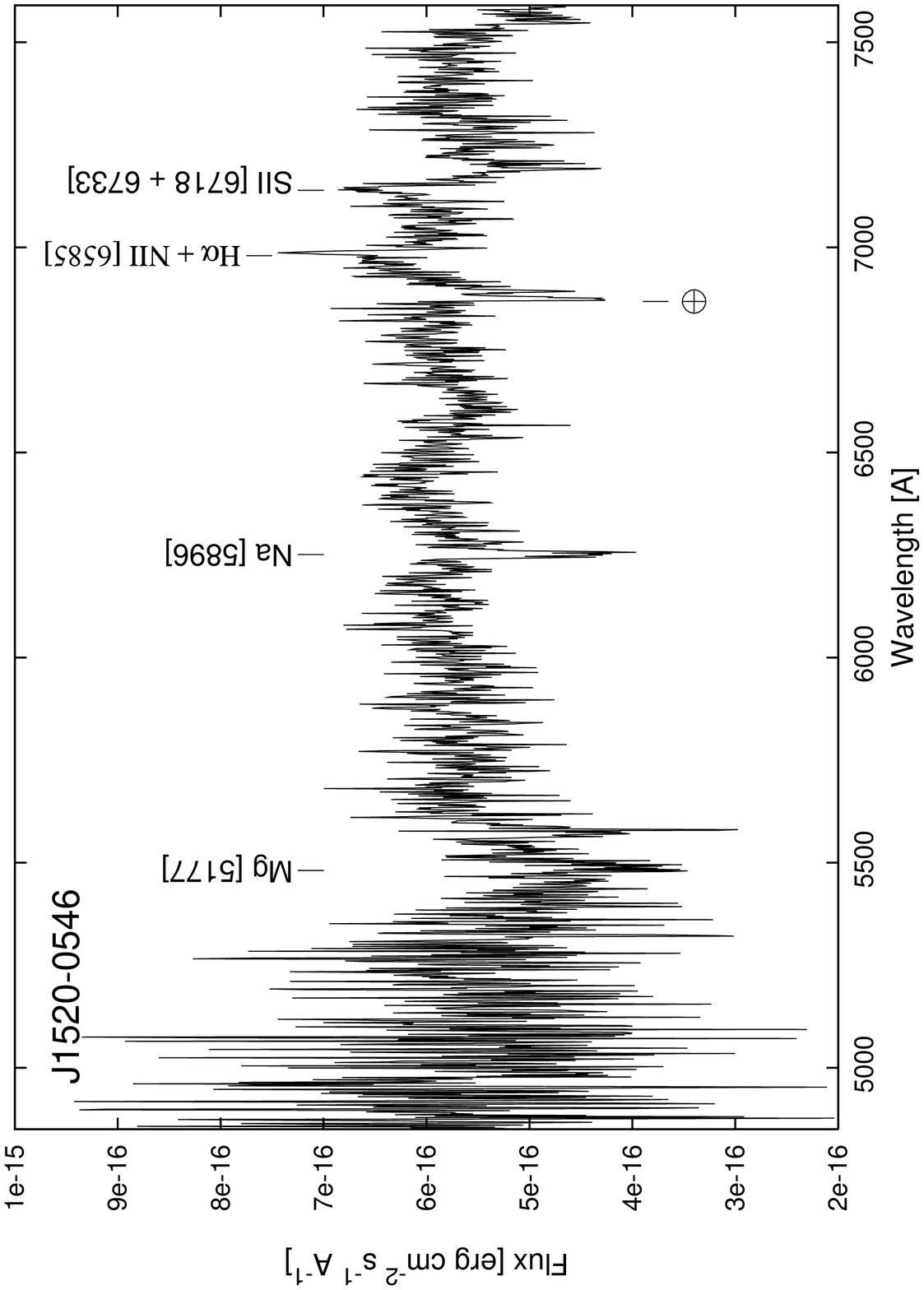}
\includegraphics{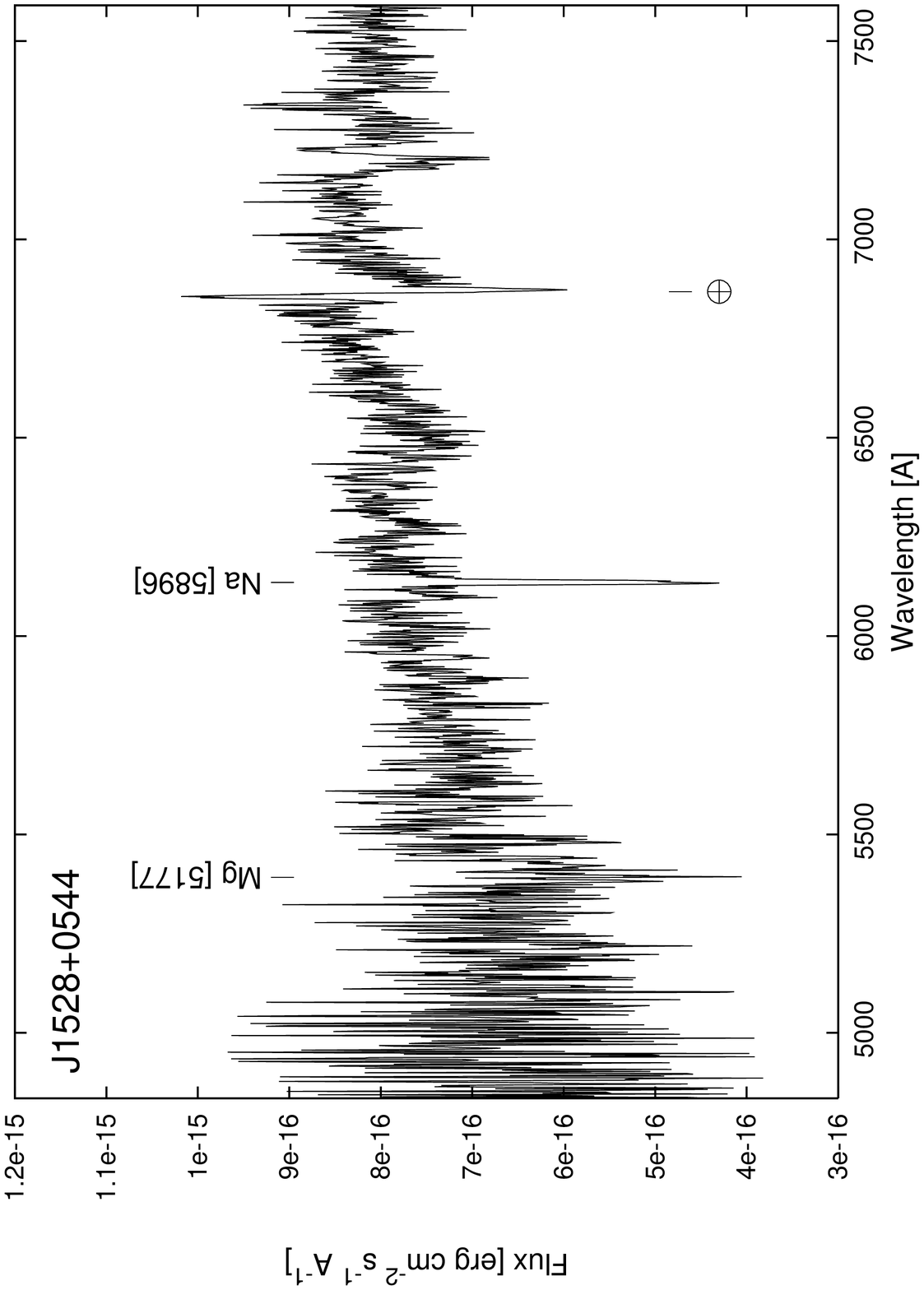}
\vspace{210mm}
\caption{The same as in Fig.~2a but for the sources: J1445$+$0932, J1445$-$0540, 
J1520$-$0546, J1528$+$0544}
\end{center}
\end{figure}

\renewcommand{\thefigure}{2d}
\begin{figure}[t]
\begin{center}
\includegraphics{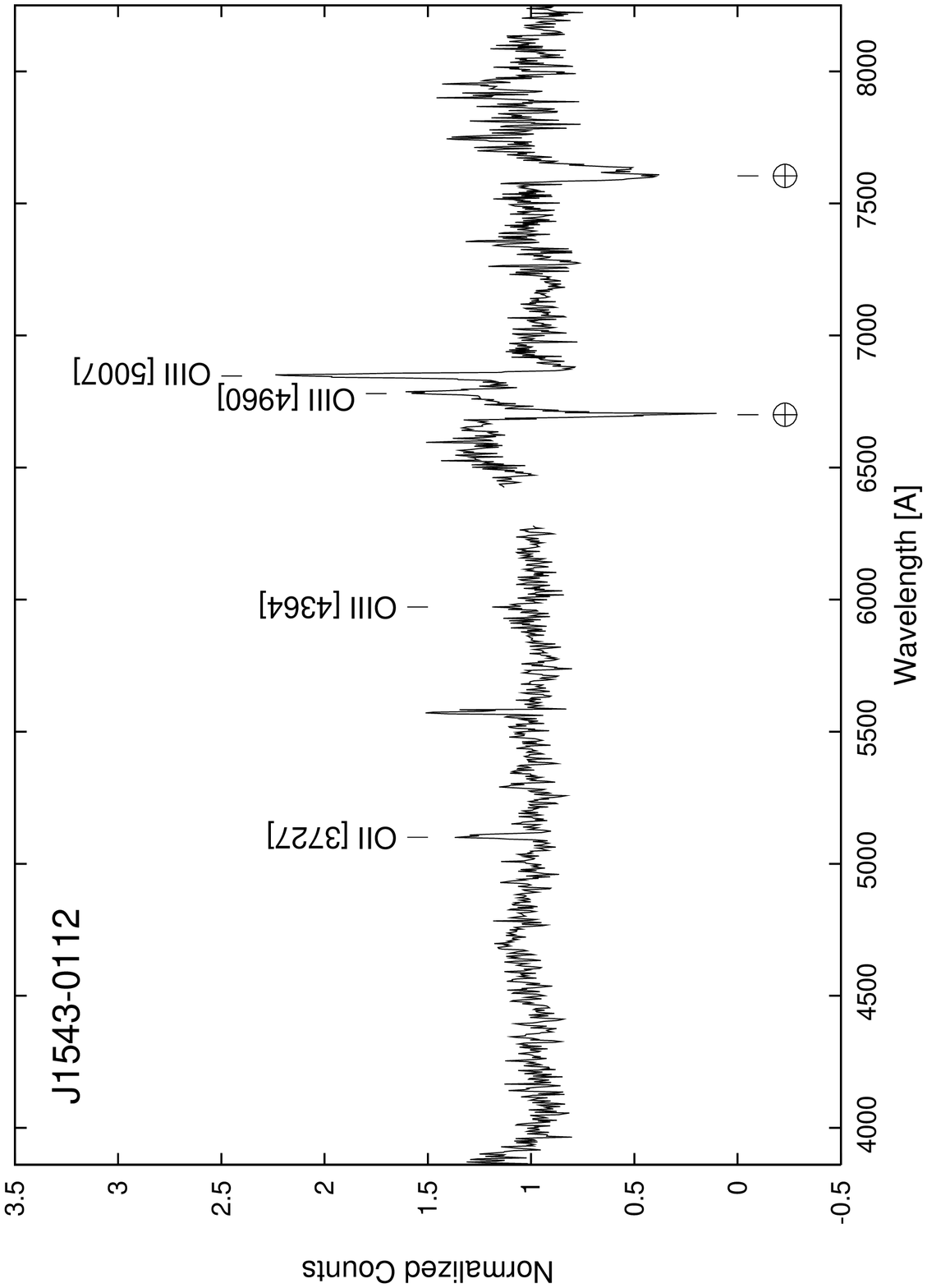}
\includegraphics{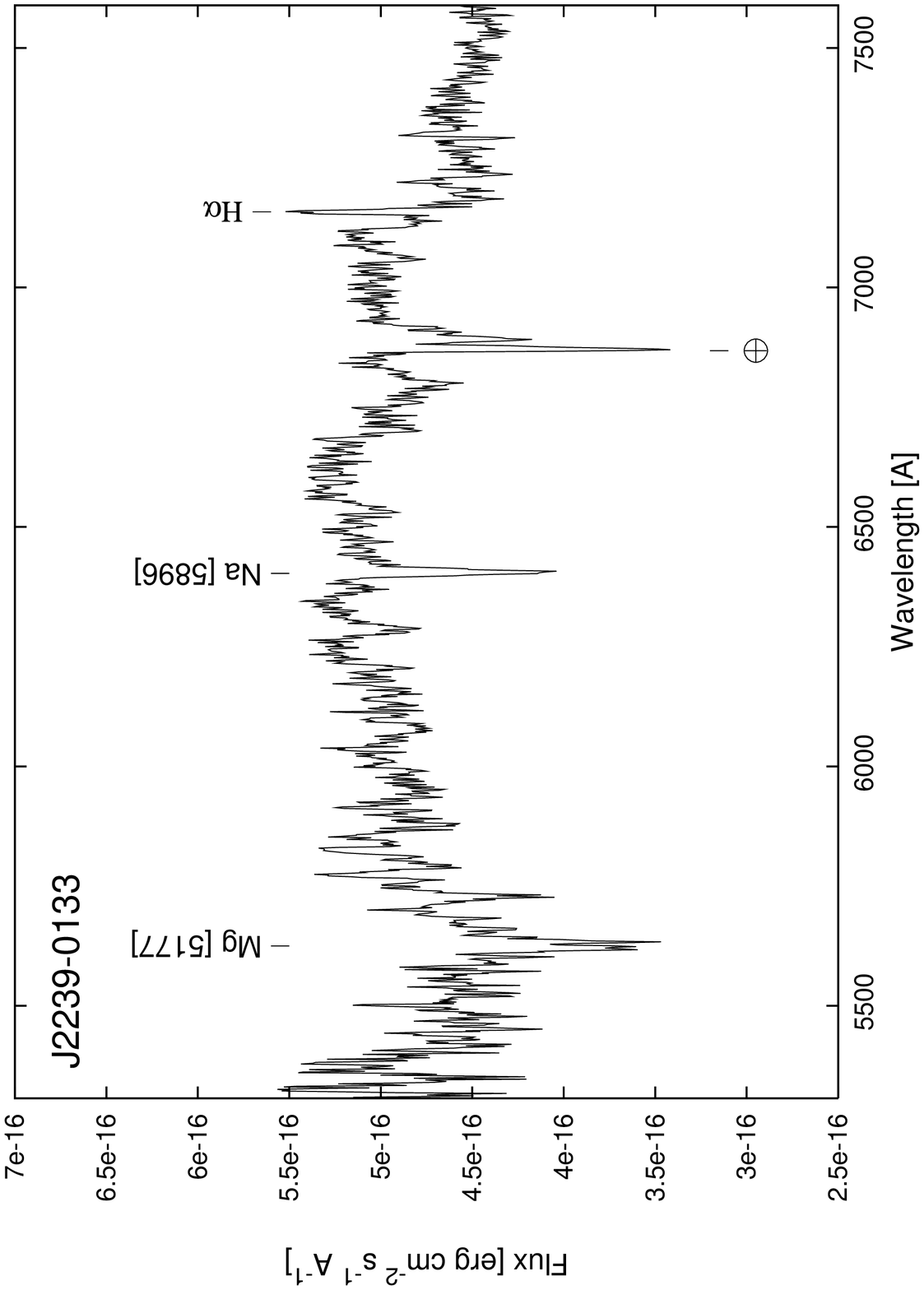}
\vspace{210mm}
\caption{The same as in Fig.~2a but for the sources: J1543$-$0112, J2239$-$0133}
\end{center}
\end{figure}

\vspace{3mm}
\noindent
\underline{5. Physical parameters}

In order to analyse and discuss our preliminary results on the IGM evolution,
we determine or estimate some global physical and geometrical parameters for the
sample sources. These parameters are defined and given in Table~6. The entries in
parentheses are the approximated values for the sources with the photometric
redshift estimate given in Table~3. The seven columns of Table~6 give:

\noindent
Col. 1: Repeated IAU name;\\
Col. 2: Logarithm of radio luminosity at the emitted frequency of 1.4 GHz calculated
with the cosmological parameters: $H_{0}$=71 km\,s$^{-1}$Mpc$^{-1}$, $\Omega_{m}$=0.27,
$\Omega_{\Lambda}$=0.73;\\
Col. 3: Projected linear size;\\
Col. 4: Average magnetic field calculated from the revised equipartition and minimum
energy formula of Beck \& Krause (2005) under assumption of a cylindrical geometry
of the extended radio emission with the base diameter set equal to the average width
of the lobes and {\bf K$_{0}$}$\equiv (E_{\rm p}/E_{\rm e})^{\alpha_{0}}$=52 for the
injection spectral index of $\alpha_{0}$=0.525;\\
Col. 5: Equipartition energy density under the same assumption;\\
Col. 6: Arm separation ratio: the separation of the farther hotspot or the lobe's
boundary from the core or the identified optical galaxy to the nearer one;\\
Col. 7: Core prominence defined as the ratio $f_{\rm c}=P_{\rm core}/(P_{\rm total}
-P_{\rm core})$, where both $P_{\rm core}$ and $P_{\rm total}$ are determined at
the emitted frequency of 1.4 GHz using the data given in Table~3, however for the
sample sources without a certain spectral index the value of 0.9 has been assumed,
and the relevant value of 0.0 has been taken for the cores.

An estimation of the magnetic field strength in extended lobes of  double radio
sources still is a controversial matter. In majority of previously published papers
the classical energy equipartition approach of Pacholczyk (1970) and its formalism of
Miley (1980) was used for this purpose. In the Pacholczyk's formula
the equipartition magnetic field strongly depends on the hardly known ratio of energy
in heavy particles to that in the electrons (+positrons) $k$; in most of the above
papers $k$=1 was assumed and the radio spectrum was integrated between a defined
frequency range, e.g. between 10 MHz and 100 GHz. In a slightly different approach
to the equipartition field calculations, the spectrum is integrated between frequencies
corresponding to a minimum and a maximum Lorentz factor for the relativistic electrons
(cf. Croston et al. 2005). To overcome the problem of uncertain value of $k$, Beck
\& Krause (2005) have proposed a revised formula for the magnetic field strength
applicable in the simple case that the number density ratio {\bf K$_{0}$} of protons
and electrons is constant which is valid in a limited range of particle energies.
They estimate that the value of this parameter is 40 $<$ {\bf K$_{0}$} $<$ 100
depending on a value of the injection spectral index $\alpha_{0}$.
The above three different estimates of average magnetic field in the lobes of selected
GRGs are compared between themselves in Konar et al. (2007).

\vspace{3mm}
\noindent
\underline{6. Discussion of the results}

\vspace{3mm}
\noindent
{\sl 6.1. Correlation between  energy density (pressure) in the lobes and  redshift} 

Since an average internal lobe pressure is proportional to the energy density, in Fig.\,3
we plot logarithm of the latter vs. the source's redshift (1+$z$) for the sample of our
GRGs, and supplement it with the corresponding data for two other samples: a limited
sample of already known GRGs as well as a sample of ``normal-sized'' FRII-type radio galaxies
taken from Machalski \& Jamrozy (2006). Because the equipartition
magnetic field calculated using the Beck \& Krause formula is larger than the Miley's
value by a factor of $\sim$3, the resulting energy densities are one order larger than
the classical values. Therefore these values in Fig.\,3 (and Table 6) are about 10 times
higher than corresponding values calculated with the Miley's formula and the lowest
values of $u_{\rm eq}$ are about 10$^{-14}$ J\,m$^{-3}$ instead of about 10$^{-15}$ J\,m$^{-3}$
like in Subrahmanyan \& Saripalli (1993) or Cotter (1998). However this difference can be
neglected if we compare the sources with similarly calculated magnetic fields.

This is clearly seen in Fig.\,3 that we find 13 GRGs with the values of $u_{\rm eq}$ lower
than the lowest values found up to now, though most of them characterize the low-redshift
sources at $z<0.2$. An exception is the source J1420$-$0545.  The solid bar
indicates an uncertainty of its redshift corresponding to the standard deviation of
about 1 mag in the distribution of absolute optical magnitudes of galaxies hosting the GRGs
(cf. Lara et al., 2001). The photometric redshift estimate of 0.32 implies a projected linear
size of over 4.8 Mpc which, if real, would be larger than the largest GRG known up to now,
i.e. 3C236!  Unfortunately, our attempts to obtain its optical spectrum either with the SALT or
to get observing time at other large telescopes were unsuccessful. The dashed straight line
marks the $u_{\rm eq}\propto$(1+$z$)$^{5}$ relation.

\begin{table}[t]
\footnotesize{
\caption{Physical parameters. The entries in parentheses concern the sample sources with the 
photometric redshift estimate}
\begin{tabular}{lccrcrrccc}
\hline
Name     & log$P_{1.4}$ & & $D$ & &$B_{\rm eq}$ & $u_{\rm eq}$$\times$10$^{-14}$ &
         & $R_{\theta}$ & $f_{\rm c}$\\
         & [W\,Hz$^{-1}$] & &[kpc] && [nT] & [J\,m$^{-3}$]\\
\hline
{\bf NGH:}\\
J0824+0140   & 24.83 & &  994 & & 0.224 &  4.67 & & 1.19 & 0.105\\
J0903+1208   &(25.40)& &(1310)& &(0.249)& (5.74)& & 1.35 & --\\ 
J0922+0919   &(25.23)& &(1580)& &(0.195)& (3.52)& & 1.52 & 0.010\\
J0925$-$0114 & 23.95 & & 1144 & & 0.129 &  1.55 & & 1.31 & ?\\
J0947$-$1338 & 25.17 & & 2258 & & 0.163 &  2.47 & & 1.13 & 0.020\\
J1005$-$1315 &(25.25)& &(1175)& &(0.240)& (5.33)& & 1.30 & --\\
J1014$-$0146 & 25.38 & &  949 & & 0.344 & 10.95 & & 1.08 & 0.086\\
J1018$-$1240 & 24.60 & &  828 & & 0.264 &  6.45 & & 1.42 & 0.089\\
J1021$-$0236 & 25.76 & & 1547 & & 0.267 &  6.61 & & 1.40 & 0.028\\
J1021+1217   & 24.77 & & 1974 & & 0.134 &  1.66 & & 1.12 & 0.124\\
J1021+0519   & 24.93 & & 2234 & & 0.137 &  1.75 & & 1.56 & 0.039\\
J1048+1108   & 25.19 & &  790 & & 0.378 & 13.27 & & 1.30 & 0.012\\
J1049$-$1308 &(25.05)& &(1085)& &(0.260)& (6.35)& & 1.40 & --\\
J1058+0812   &(25.24)& &(1730)& &(0.168)& (2.62)& & 1.25 & --\\
J1101$-$1053 &(24.74)& &(2200)& &(0.115)& (1.24)& & 1.24 & --\\
J1108+0202   & 25.13 & & 1497 & & 0.213 &  4.23 & & 1.22 & --\\
J1126$-$0042 & 25.76 & & 1132 & & 0.340 & 10.76 & & 1.77 & 0.008\\
J1130$-$1320 & 27.29 & & 2033 & & 0.419 & 16.07 & & 1.04 & --\\
J1213$-$0500 & 23.82 & &  849 & & 0.144 &  1.92 & & 1.02 & --\\
J1253$-$0139 &(25.43)& & (868)& &(0.378)&(13.27)& & 1.25 & --\\
J1328$-$0129 & 25.31 & &  841 & & 0.352 & 11.52 & & 1.00 & 0.045\\
J1328$-$0307 & 24.56 & & 1290 & & 0.183 &  3.11 & & 1.16 & 0.041\\
J1334$-$1009 & 25.50 & & 1322 & & 0.322 &  9.64 & & 1.11 & 0.053\\
J1354$-$0705 &(25.00)& & (750)& &(0.325)& (9.82)& & 1.60 & 0.020\\
J1411+0619   & 25.67 & & 1547 & & 0.232 &  5.00 & & 1.20 & ?\\
J1420$-$0545 &(25.44)& &(4830)& &(0.095)& (0.83)& & 1.08 & 0.024\\
J1445+0932   & 24.49 & &  543 & & 0.355 & 11.70 & & 1.05 & 0.043\\
J1445$-$0540 & 26.02 & & 2124 & & 0.227 &  4.76 & & 1.24 & 0.042\\
J1457$-$0613 & 25.58 & &  715 & & 0.515 & 24.65 & & 1.30 & --\\
J1459$-$0432 &(25.14)& &(1660)& &(0.187)& (3.25)& & 1.42 & 0.010\\ 
J1520$-$0546 & 23.73 & & 1525 & & 0.091 &  0.77 & & 1.06 & 0.192\\
J1528+0544   & 23.95 & &  643 & & 0.222 &  4.56 & & 1.39 & 0.086\\
J1540$-$0127 & 25.09 & &  759 & & 0.363 & 12.30 & & ?    & 0.044\\
J1543$-$0112 & 25.70 & &  887 & & 0.393 & 14.37 & & 1.30 & 0.003\\
{\bf SGH:}\\
J2234$-$0224 & 25.94 & & 1266 & & 0.279 &  7.24 & & 1.45 & 0.060\\
J2239$-$0133 & 24.39 & & 1960 & & 0.119 &  1.31 & & 2.22 & 0.013\\
J2345$-$0449 & 24.35 & & 1454 & & 0.158 &  2.32 & & 1.22 & 0.021\\
J0010$-$1108 & 23.90 & &  795 & & 0.173 &  2.79 & & 1.35 & 0.100\\
J0037+0027   &(25.90)& &(1650)& &(0.250)& (5.85)& & 1.25 & 0.011\\
J0039$-$1300 & 24.73 & & 1228 & & 0.208 &  4.00 & & 1.14 & --\\
J0042$-$0613 & 25.64 & &  852 & & 0.310 &  8.92 & & ?    & --\\
J0129$-$0758 & 24.39 & & 1300 & & 0.164 &  2.51 & & 1.10 & --\\
J0134$-$0107 & 24.63 & & 1209 & & 0.166 &  2.56 & & ?    & --\\   
J0202$-$0939 &(26.28)& & (785)& &(0.640)& (38.0)& & 1.41 & 0.001\\
J0300$-$0728 & 25.67 & & 1806 & & 0.183 &  3.12 & & 1.43 & 0.017\\
J0313$-$0632 & 25.93 & & 1024 & & 0.390 & 14.10 & & 1.55 & --\\     
\hline
\end{tabular}}
\end{table}

\renewcommand{\thefigure}{3} 
\begin{figure}[h]
\begin{center}
\includegraphics{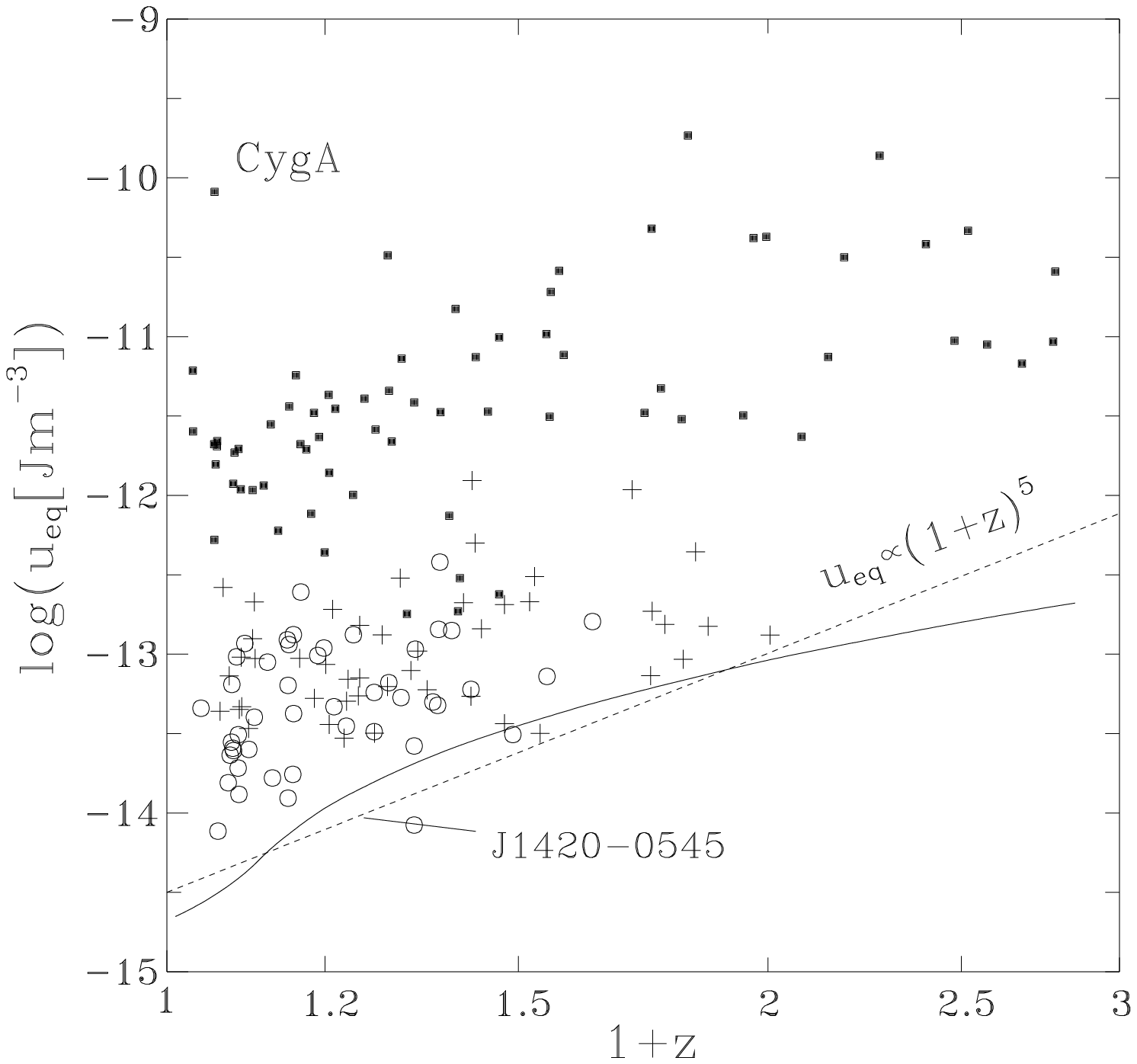}
\vspace{145mm}
\caption{The equipartition energy density of radio sources as a function of
redshift. Selected ``normal-sized'' sources are marked with the full squares.
A number of already known GRGs are marked with crosses, while the newly detected
GRGs (given in Table 5) are marked with the open circles. The dashed line follows
the proportionality $p\propto (1+z)^{5}$. An uncertainty of the redshift
estimate for the sample GRG J1420$-$0545 is marked with the full line, cf. the
text. The solid curve indicates a relation $u_{eq}\propto(P/V)^{4/7}$, cf. the text}
\end{center}
\end{figure}

Anyhow the data presented in this paper confirm the known strong increase of the
equipartition energy density with increasing redshift (cf. Subrahmanyan \& Saripalli
1993; Cotter 1998). To investigate how the observed $u_{\rm eq}$ vs. (1+$z$)
correlation reflects an expected change in the environment of GRGs one has to
check how the selection effects affect the diagram shown in Fig.~3. According to the
classical formula, the equipartition energy density is related to the total power
and the volume of a radio source by $u_{\rm eq}\propto(P/V)^{4/7}$. Since all the
sources plotted in the diagram belong to the flux-density limited samples, therefore
sources at higher redshift have, on average, higher radio powers and thus higher
energy density at equal source volumes. Moreover, we can expect that an average volume
of sources decreases with redshift as does their average linear size. Following
Schoenmakers et al. (2000) we calculate the expected $u_{\rm eq}$ as a function of
(1+$z$) for sources with flux density of 45 mJy, i.e. a mean value of the limiting
fluxes in the NGH and SGH subsamples, and a volume corresponding to cylinder length
of 1 Mpc and its diameter of 167 kpc. The resulting relation is shown with the solid
curve whose shape simply indicates the log\,$P$--log(1+$z$) relation for a constant
flux density and the radio spectral index of unity. 

In our previous paper (Machalski \& Jamrozy 2006) we showed that performing a
statistical test for correlations between two variables in the presence of one or
two other variables one can examine whether a residual correlation between the above
two variables (parameters) is still present when the third (or the third and fourth)
is (are) held constant. In particular, we showed that the equipartition energy density
calculated for the sample radio sources, but transformed to a reference total power
and a reference linear size, practically did not correlate with redshift. It seems 
that supplementing their data (crosses and full dots in Fig.~3) with further
GRGs investigated in this paper (open circles in Fig.~3) do not change the situation
significantly. 

\vspace{3mm}
\noindent
{\sl 6.2. Morphology of GRGs vs. the environment}

Projected linear sizes of the largest FRII-type radio sources, considerably exceeding
those predicted (at different redshifts) by the very early dynamical model of 
Gopal-Krishna \& Wiita (1987), already suggested that the IGM may be inhomogeneous
and large-scale density fluctuations, especially at high redshifts, can play a
significant role in the time evolution of those sources.

The unified scheme for radio sources with active galactic nuclei (AGN) (e.g. Barthel
1989; Gopal-Krishna, Kulkarni \& Mangalan 1994; Urry \& Padovani, 1995) predicts how
the appearance of sources evolving in a homogeneous environment depends on orientation
of the jets' axis towards the observer's line of sight. This appearance includes the
arm separation ratio  $R_{\theta}$, the core prominence parameter, $f_{\rm c}$ (both
defined in Section~5), and the flux density ratio in radiation from the opposite
lobes (not analysed in this paper). Within this scheme, the effects of relativistic
beaming cause that the sources at small inclination angles should have more prominent
cores and higher asymmetry ratios in the arm length and flux density compared with
those at larger angles.

The evident lack of a statistical correlation between $R_{\theta}$ and $f_{\rm c}$
in our sample (the correlation coefficient of $-$0.38 suggesting even an anticorrelation)
strongly supports a common conviction that the inclination angle of majority of GRGs
is close to 90$^{\rm o}$. Therefore, because the lobes of GRGs lie well beyond the
boundary of the parent optical galaxies, their symmetry/asymmetry  parameters may
reflect eventual environmental heterogeneity. Ishwara-Chandra \& Saikia (1999) and
Schoenmakers et al. (2000) noted that the GRGs tended to be
marginally more asymmetric than smaller sources of a comparable radio luminosity.
For example, Ishwara-Chandra \& Saikia noted that the median value of the arm separation
ratio in their sample of GRGs is about 1.39 compared with about 1.19 for 3CR galaxies
of similar luminosity but smaller sizes. According to the unified scheme, the expected
arm separation ratio for a source inclined at $i>45^{o}$ to the line of sight is only
about 1.15 for the jet head advance  speed of $\sim$0.1 (cf. Scheuer 1995). As the
inclination angles of GRGs very likely are close to 90$^{\rm o}$, the arm asymmetry ratio
(frequently much higher than one) should be related to the environmental properties
rather than to an orientation towards the observer.

The radio data available for the GRGs investigated in this paper allow us to study
a distribution of $R_{\theta}$, and (within a limited range)  $f_{\rm c}$. The median
value of 1.25$\pm$0.03
in the distribution of $R_{\theta}$ in our sample supports the Ishwara-Chandra \& Saikia's
conclusion that the apparent arm asymmetry in the GRGs {\sl ``could possibly be caused
by interaction of the energy-carrying beams with cluster-sized density gradients far
from the parent galaxy''}. The presented analysis of the sample of faint low-luminosity
GRGs whose lobes lie at least at 0.5--1 Mpc from the parent AGN strongly suggests that
these sources, after probing a denser central-halo environment, might further evolve in
exceptionally low-density intergalactic medium to some degree being ``voids'' in the hot IGM.
Therefore, search for low-luminosity, high-redshift GRGs and the analysis of their physical
and structural parameters, which is the aim of our SALT project, should help to study
these density fluctuations and find their relation to the theoretically predicted
cosmological evolution of the IGM density due to the expansion of the Universe.

\vspace{2mm}
\noindent
Bearing in mind the above we can conclude as follows:

-- If the internal pressure in diffuse lobes really counterbalances the IGM pressure,
the data collected till now do not provide a certain observational evidence for
the strong cosmological evolution of the latter in the form $p_{\rm IGM}(z)\propto (1+z)^{5}$.
However, if the lobes of the most extended and diffused sources are still
overpressured in respect to the IGM, especially at low redshifts, any conclusion about
the properties of the IGM will not be reliable.

-- Deduced low internal pressure in the lobes of GRGs, their observed morphology and
the arm asymmetry strongly suggest an existence of deep gradiants in the IGM density with
the scale of a few Mpc. However, because an internal pressure
in the lobes is proportional to their avarage energy density, a value of this pressure
depends of how the magnetic field strength is determined. On the basis of the
{\sl Chandra} and {\sl XXM-Newton} data on X-ray emission from the lobes of classical
double radio sources, Croston et al. (2005) found that ``the measured X-ray flux can be
attributed to inverse-Compton scattering of the cosmic microwave background radiation
with magnetic field strengths in the lobes between 0.3$B_{\rm eq}$ and 1.3$B_{\rm eq}$,
where the value $B_{\rm eq}$ corresponds to equipartition between the electrons and
magnetic field, assuming a filling factor of unity''. Therefore the equipartition
magnetic field, energy density, and thus internal lobe-pressure estimates, at least
in ``normal-sized'' sources are likely close to their true values. Unfortunately,
X-ray data are not yet available for the lobes of GRGs, in particular for those of
low radio luminosity.

-- There is still a hope that detection of the most distant and low-luminosity GRGs,
which is the aim of our SALT project, can help in an observational confinement
on the hypothesis of cosmological evolution of the IGM. The most promising is our SGH
subsample consisting of many GRG candidates optically not identified on the DSS frames,
thus with apparent magnitudes of $R>20.5$ mag.

\vspace{4mm}
\noindent
{\bf Acknowledgements} 
The authors thank Dr Alexei Kniazev and the remaining SALT staff for their 
effort in making the PFIS observations. D. K.-W. acknowledges the hospitality of 
the SAAO authorities and Dr Stephen Potter for helping in the observations. 
We also also thank Micha{\l} Siwak for the Mt. Suhora observations. This project 
was supported by MNiSW with funding for scientific research in the years
2005--2007 under contract No. 0425/PO3/2005/29. Support from SALT International 
Network grant No. 76/E-60/SPB/MSN/P-03/DWM 35/2005-2007 is also acknowledged.

\end{document}